\documentclass[preprint,preprintnumbers,amsmath,amssymb,floatfix,nofootinbib]{revtex4}
\usepackage{epsf, array, color}
\usepackage{graphicx}
\usepackage{subfigure}
\usepackage{slashed}

\usepackage{bbold}
\usepackage{amsmath}
\usepackage{dcolumn}% Align table columns on decimal point

\usepackage{epsfig}
\usepackage{amssymb,amsmath}

\usepackage{ulem} % to use a strikeout/strikethrough font

\newcommand{\gsim}{\gtrsim}

\newcommand{\bea}{\begin{eqnarray}}
\newcommand{\eea}{\end{eqnarray}}
\newcommand{\nn}{\nonumber \\}
\newcommand{\lag}{\ensuremath{{\cal L}}}

\def\beq{\begin{equation}}
\def\eeq{\end{equation}}

\begin{document}

\begin{flushleft} 
%{\it preprint } 
\end{flushleft} 
\begin{center} 
\Large Vector-like Fermions and Higgs Effective Field Theory Revisited
\end{center} 
\numberwithin{equation}{section}
\renewcommand{\theequation}{\arabic{section}.\arabic{equation}} 

\author{Chien-Yi Chen}
\affiliation{
 Department of Physics and Astronomy,
University of Victoria, Victoria, BC~ V8P~ 5C2, Canada}
\affiliation{Perimeter Institute for Theoretical Physics, Waterloo, ON~ N2J~ 2W9, Canada}

\author{S. Dawson}
\affiliation{Department of Physics,
 Brookhaven National Laboratory, Upton, N.Y., 11973 U.S.A.}
 
 \author{Elisabetta Furlan}
\affiliation{Institute for Theoretical Physics, ETH, 8093~Zurich, Switzerland}
 % \vspace*{1cm}}

\date{\today}

\begin{abstract}
Heavy vector-like quarks (VLQs) appear in many models of beyond the Standard Model physics. Direct
experimental searches require these new quarks to be heavy, $\gsim 800-1000$~GeV. 
We perform a global fit of the parameters of simple VLQ models in minimal representations 
of $SU(2)_L$ to precision data and Higgs rates. An interesting connection between anomalous 
$Z b {\overline {b}}$ interactions and Higgs physics in VLQ models is discussed.  
Finally, we present our analysis  in an 
effective field theory (EFT) framework and show  that the parameters of VLQ models are already
highly constrained. 
Exact and approximate analytical formulas for the $S$ and $T$ parameters 
in the VLQ models we consider are posted at 
\href{$quark.phy.bnl.gov\backslash Digital\_Data\_Archive\backslash dawson\backslash vlq\_17$}{https://quark.phy.bnl.gov/Digital_Data_Archive/dawson/vlq_17/} 
as Mathematica files.

\end{abstract}

\maketitle

%%%%%%%%%%%%%%%%%%%%%%%%%%%%%%%%%%%%%%%%%%%%%%%%%%%%%%%%%%%%%%%%%%%%%%%
\newpage
%%%%%
\section{Introduction}
The Standard Model (SM) has been remarkably successful at explaining both precision 
measurements and LHC data and so the possibilities for heavy, as yet unobserved 
particles are highly restricted by the experimental results. Here,
we focus on new heavy quarks and their impact on electroweak
scale physics. Heavy SM-like chiral fermions are excluded by the measured Higgs production 
rates~\cite{Anastasiou:2011qw,Anastasiou:2016cez}. Therefore, we consider heavy vector-like 
quarks (VLQs), 
which are typically compatible with Higgs measurements. Motivated by the excellent 
agreement of Higgs measurements with SM predictions~\cite{ATLAS-CONF-2015-044}, 
we assume that the observed Higgs boson is part of an $SU(2)_L$ doublet, $H$, and 
consider VLQs which can couple to $H$. This class of VLQs occurs in many 
composite Higgs models~\cite{Anastasiou:2009rv,Contino:2010rs,Contino:2006qr,Agashe:2005dk,Gillioz:2012se,Panico:2015jxa} and little Higgs 
models~\cite{ArkaniHamed:2002qy,Han:2005ru,Hubisz:2004ft} and hence is well 
motivated phenomenologically. The phenomenology of VLQs has been considered in some detail 
in the literature~\cite{Lavoura:1992np,Burgess:1993vc,delAguila:1998tp,
AguilarSaavedra:2002kr,Cacciapaglia:2010vn,Dawson:2012di,
Aguilar-Saavedra:2013qpa,Ellis:2014dza,Angelescu:2015kga,Fajfer:2013wca,
Alok:2015iha,Alok:2014yua} and
direct experimental searches~\cite{ATLAS-CONF-2016-101,ATLAS-CONF-2016-102,
ATLAS-CONF-2016-013,Aad:2016shx,Aad:2016qpo,Aad:2015voa, 
Aad:2015kqa,Aad:2015mba,Khachatryan:2015gza,Khachatryan:2015axa,
Chatrchyan:2013wfa,Chatrchyan:2013uxa, ATLAS-CONF-2017-015} 
require them to be heavy, with $M \gsim {\cal{O}}(800-1000)$~GeV. 

We update previous fits~\cite{Dawson:2012di,Aguilar-Saavedra:2013qpa,Ellis:2014dza} 
to the parameters of VLQ models by performing a joint fit to the oblique parameters
and asymmetries in the $b$ quark sector.  
The study is extended to include
restrictions from Higgs coupling measurements 
with interesting results found in models containing a $B$ VLQ.

We briefly review the set-up of the VLQ models that we study in 
Section~\ref{vlqbas}. Section~\ref{vlqob} 
reviews the contributions of VLQs to the oblique parameters and the 
$Zb{\overline{b}}$ coupling.
% and discusses some numerical stability issues.  
We find that in some regions of parameter space the leading contributions 
to the oblique parameters can be quite small even with significant mass 
splittings between the VLQ multiplet members, due to numerical cancellations.  
We discuss the effects of these regions on the global fits to VLQ parameters. 
Section~\ref{globsec} contains numerical
fits and we present some conclusions in Section~\ref{consec}. Appendix~\ref{VLTapp} 
contains a pedagogical description of the triplet models, which  should be useful for model builders. 
The connection between our results in the
full VLQ theories and in an EFT approach for heavy VLQ masses is given  in Appendix~\ref{eftapp}. 
Exact and approximate analytical formulas for the oblique parameters in the various models
can be found at \href{$quark.phy.bnl.gov\backslash Digital\_Data\_Archive\backslash dawson\backslash vlq\_17$}{https://quark.phy.bnl.gov/Digital_Data_Archive/dawson/vlq_17/}.

\section{Vector-Like Quark Basics}
\label{vlqbas}
\subsection{Basics}
In this section, we introduce our notation for VLQs. We consider the case where 
the VLQs interact only with the third
generation quarks, since mixing with the first two generations is highly restricted by kaon
and other low energy physics measurements~\cite{AguilarSaavedra:2002kr}.
We indicate the SM weak eigenstate quarks as,
\begin{equation}
\psi_L^0=\left(\begin{matrix}
t_L^0\\
b_L^0\end{matrix}\right), \quad t_R^0, b_R^0\, ,
\end{equation}
and the Higgs doublet as,
\begin{equation}
H = \left(\begin{matrix}\phi^+\\ \phi^0\end{matrix}\right) \,,
\end{equation}
with $\phi^0={v+h \over \sqrt{2}}$. The SM Yukawa couplings are,
\begin{equation}
-L_{Y, SM}=\lambda_t{\overline \psi}_L^0 {\tilde H}t_R^0 +
\lambda_b{\overline \psi}_L^0 { H}b_R^0+h.c.\, ,
\label{ysm}
\end{equation}
where ${\tilde H}=i\sigma_2 H^{*}$. 

The models we consider have vector-like quarks in the $SU(2)_L$ representations,
\begin{eqnarray}
&{\rm Singlets:}
	& T_s^0, \quad B_s^0 \, ;\nonumber \\
&{\rm Doublets:}
	& \psi_{XT}^0=(X_d^0,T_d^0), \nonumber \\
	&&\psi_{TB}^0=(T_d^0, B_d^0), \nonumber \\
	&&\psi_{BY}^0= (B_d^0, Y_d^0) \,;\nonumber \\
&{\rm Triplets:} & \rho_{XTB}^0=(X_t^0, T_t^0, B_t^0), \nonumber \\
	&&\rho_{TBY}^0=(T_t^0, B_t^0, Y_t^0)\, .
\end{eqnarray}

This is a complete set of VLQ representations that have renormalizable couplings to the  SM Higgs doublet.
The quarks have electric charge $Q_T={2\over 3},~Q_B=-{1\over 3},~Q_X={5\over 3}$,
and $Q_Y=-{4\over 3}$.
If there is only one VLQ representation, it is simple to write the most general CP conserving  couplings
between the SM fermions, the VLQs, and the Higgs boson,\footnote{This can be straightforwardly generalized
to models with more than one VLQ representation~\cite{Cacciapaglia:2015ixa,Angelescu:2015kga,Ellis:2014dza}.}
\begin{eqnarray}
{\rm Singlets:}&-L_{T_s}&=
%\lambda_1{\overline \psi} _L^0 {\tilde H}(T_s ^0)_{_R}
%+M_{T_s}( {\overline T}^0_s)_{_L}(T^0_s)_{_R}+h.c.\nonumber \\
 \lambda_1{\overline \psi} _L^0 {\tilde H}T_{(s),R}^0
+M_{T_s} {\overline T}^0_{(s),L}T^0_{_(s),R}+h.c.\nonumber \\
&-L_{B_s}&=
 \lambda_2{\overline \psi} _L^0 { H}B_{(s),R}^0
+M_{B_s} {\overline B}^0_{(s),L}B^0_{(s),R}+h.c.\nonumber \\
{\rm Doublets:}& -L_{XT}&=\lambda_3{\overline \psi}_{(XT),L}^0 H t_R^0
+M_{XT}{\overline \psi}_{(XT),L}^0\psi_{(XT),R}^0+h.c.\nonumber \\
&-L_{TB}&=\lambda_{4}{\overline \psi}_{(TB),L}^0 {\tilde H}t_R^0
+ \lambda_{5}{\overline \psi} _{(TB),L}^0 { H}b_{R}^0
+M_{TB} {\overline \psi}_{(TB),L}^0\psi_{(TB),R}^0+h.c.\nonumber \\
&-L_{BY}&= \lambda_{6}{\overline \psi}_{(BY),L}^0 {\tilde H} b_R^0
+M_{BY}{\overline \psi}_{(BY),L}^0\psi_{(BY),R}^0+h.c.\nonumber \\
%Triplets: & -L_{XTB}&=\lambda_{7}{\overline \rho}_{XTB}^0 
%{\tilde H}^\dagger\sigma \psi_L^0 
%+M_{XTB} {\overline \rho}_{XTB}^0\rho_{XTB}^0+h.c. \nonumber 
%\\
%& -L_{TBY} &=\lambda_{8}{\overline \rho}_{TBY}^0 {\tilde H}\sigma \psi_L^0 +
%M_{TBY}{\overline \rho}_{TBY}^0\rho_{TBY}^0+h.c.  \nonumber \\
%&-L_{B_s}&=
% \lambda_2{\overline \psi} _L^0 { H}B_{sR}^0
%+M_{B_s} {\overline B}^0_{sL}B^0_{sR}+h.c.\nonumber \\
%Doublets: & -L_{XT}&=\lambda_3{\overline \psi}_{(XT),L}^0 H t_R^0
%+M_{XT}{\overline \psi}_{(XT),L}^0\psi_{(XT),R}^0+h.c.\nonumber \\
%&-L_{TB}&=\lambda_{4}{\overline \psi}_{(TB),L}^0 {\tilde H}t_R^0
%+ \lambda_{5}{\overline \psi} _{(TB),L}^0 { H}b_{R}^0
%+M_{TB} {\overline \psi}_{(TB),L}^0\psi_{(TB),R}^0+h.c.\nonumber \\
%&-L_{BY}&= \lambda_{6}{\overline \psi}_{(BY),L}^0 {\tilde H} b_R^0
%+M_{BY}{\overline \psi}_{(BY),L}^0\psi_{(BY),R}^0+h.c.\nonumber \\
%Triplets: & -L_{XTB}&=\lambda_{7}{\overline \rho}_{XTB}^0 
%{\tilde H}^\dagger\sigma \psi_L^0 
%+M_{XTB} {\overline \rho}_{XTB}^0\rho_{XTB}^0+h.c. \nonumber 
%\\
%& -L_{TBY} &=\lambda_{8}{\overline \rho}_{TBY}^0 {\tilde H}\sigma \psi_L^0 +
%M_{TBY}{\overline \rho}_{TBY}^0\rho_{TBY}^0+h.c.  \nonumber \\
%
%
%
%Betta's lagrangians
{\rm Triplets:}& -L_{XTB}&= \lambda_7 \bar{\psi}_L^0 \sigma^a \rho^{0,a}_{XTB} \tilde{H} 
	+ M_{XTB} \bar{\rho}^0_{XTB} \rho^0_{XTB} +h.c. \nonumber \\
& -L_{TBY} &=\lambda_8 \bar{\psi}_L \sigma^a \rho^{0,a}_{TBY} H 
	+ M_{TBY} \bar{\rho}^0_{TBY} \rho^0_{TBY} +h.c. 
\label{genpot}
\end{eqnarray}
Note that we do not include mixing between SM fermions and VLQs with
identical quantum numbers since these terms can be rotated away by redefinitions of the fields. 
The singlet and doublet models have been extensively discussed in the
literature~\cite{
Aguilar-Saavedra:2013qpa,Ellis:2014dza,Dawson:2012di,Burgess:1993vc,Lavoura:1992np,
Cacciapaglia:2010vn,Angelescu:2015kga,delAguila:1998tp,AguilarSaavedra:2002kr}, and we include
a useful discussion of the details of the triplet model in Appendix A. 

The gauge eigenstate
fields can be written in general as,
\begin{eqnarray}
{\cal T}_{L,R}^{0}=&\left(\begin{matrix}
t_{L,R}^0\\T_{L,R}^0\end{matrix}\right)\qquad 
{\cal B}_{L,R}^{0}=&\left(\begin{matrix} b_{L,R}^0\\B_{L,R}^0\end{matrix}\right)%\nonumber \\
%\chi_R^{t0}=&\left(\begin{matrix}
%t_R^0\\T_R^0\end{matrix}\right)\qquad 
%\chi_R^{b0}=&\left(\begin{matrix} b_R^0\\B_R^0\end{matrix}\right) 
\,
\label{eq:gauge_eigenst_fields}
\end{eqnarray}
where $T^0=T^0_s, T^0_d$ or $T^0_t$ and $B^0=B^0_s, B^0_d$ or $B^0_t$ (the $X$ and $Y$
fields do not mix with the other fermions and are therefore also mass eigenstates).
The terms contributing to the mass matrices are found from
Eq.~\ref{genpot} and we write them as,
\begin{equation}
%-L_M={\overline \chi}_L^{t0}M^t\chi_R^{t0}\ +{\overline \chi}_L^{b0}M^b \chi_R^{b0}
%+M_Y {\overline Y}_{t,L}Y_{t,R}+M_X {\overline X}_{t,L}X_{t,R}
-L_M={\overline {\cal T}}_L^{0}M^t{\cal T}_R^{0}\ +{\overline {\cal B}}_L^{0}M^b {\cal B}_R^{0}
+M_Y {\overline Y}_{t,L}Y_{t,R}+M_X {\overline X}_{t,L}X_{t,R}
\, .
\end{equation}
We denote the mass eigenstate fields as $(t,T)$ and $(b,B)$ and they are found through bi-unitary transformations,
\begin{eqnarray}
%\chi_{L,R}^t&=&\left(\begin{matrix} t_{L,R}\\T_{L,R}\end{matrix}\right)=V_{L,R}^t \left(\begin{matrix} t^0_{L,R}\\T^0_{L,R}\end{matrix}\right)\quad\nonumber \\
%\chi_L^b&=&\left(\begin{matrix} b_{L,R}\\B_{L,R}\end{matrix}\right)=V_{L,R}^b \left(\begin{matrix} b^0_{L,R}\\B^0_{L,R}\end{matrix}\right)
{\cal T}_{L,R}&=&\left(\begin{matrix} t_{L,R}\\T_{L,R}\end{matrix}\right)=V_{L,R}^t \left(\begin{matrix} t^0_{L,R}\\T^0_{L,R}\end{matrix}\right)\nonumber \\
{\cal B}_{L,R}&=&\left(\begin{matrix} b_{L,R}\\B_{L,R}\end{matrix}\right)=V_{L,R}^b \left(\begin{matrix} b^0_{L,R}\\B^0_{L,R}\end{matrix}\right)
\, ,
\end{eqnarray}
where
\begin{equation}
V_{L,R}^{t,b}=\left(
\begin{matrix}
\cos\theta_{L,R}^{t,b}& -\sin\theta_{L,R}^{t,b}\\
\sin\theta_{L,R}^{t,b} &\cos\theta_{L,R}^{t,b}\end{matrix}
\right)\, ,
\label{eq:rotation_matrices}
\end{equation}
For simplicity of notation we abbreviate $\cos\theta_L^t\equiv c_L^t$, etc. 
Through these rotations we obtain the diagonal mass matrices 
\begin{equation}
M^t_{diag}=V_L^t M^t (V_R^{t})^\dagger=\left(\begin{matrix}m_t & 0 \\0 &M_T\end{matrix}\right)
\quad , \quad
M^b_{diag}=V_L^b M^b (V_R^b)^\dagger=\left(\begin{matrix}m_b & 0 \\0 &M_B\end{matrix}\right)\, .
\end{equation}

There are relationships between the angles and mass eigenstates that
depend on the representation (see for example~\cite{Aguilar-Saavedra:2013qpa}),
 \begin{eqnarray}
{\rm Doublets} 
&(XT): &M_X^2=M_T^2 (c_R^t)^{2}+m_t^2 (s_R^{t})^{2}\nonumber \\
&(TB): &M_T^2 (c_R^{t})^{2}+m_t^2 (s_R^{t})^{2}=M_B^2 (c_R^{b})^{2}+m_b^2 (s_R^{b})^{2}
\nonumber \\
&(BY):&M_Y^2=M_B^2 (c_R^b)^{2}+m_b^2 (s_R^b)^{2}\nonumber \\
{\rm Triplets}& (XTB): & M_X^2=M_T^2 (c_L^t)^{2}+m_t^2 (s_L^t)^{2}\nonumber \\
 %&& M_T^2 (c_L^t)^{2}+m_t^2 (s_L^t)^{2}=M_B^2 (c_L^b)^{2} + m_b^2
 %(s_L^b)^{2}\nonumber \\
&& \phantom{M_X^2} =M_B^2 (c_L^b)^{2} + m_b^2(s_L^b)^{2}\nonumber \\
  &&\sin(2\theta_L^b)= \sqrt{2}{M_T^2-m_t^2\over  (M_B^2-m_b^2)}\sin(2\theta_L^t)\nonumber \\
  &(TBY):& M_Y^2=M_B ^2 (c_L^b)^2+m_b^2 (s_L^b)^{2}\nonumber \\
 %&& M_T^2 (c_L^t)^{2}+m_t^2 (s_L^t)^{2}=M_B^2 (c_L^b)^{2}+m_b^2
 %(s_L^b)^{2} \nonumber \\
  &&\phantom{M_Y^2}=M_T^2 (c_L^t)^{2}+m_t^2 (s_L^t)^{2} \nonumber\\
 &&\sin(2\theta_L^b)= {M_T^2-m_t^2\over \sqrt{2} (M_B^2-m_b^2)}\sin(2\theta_L^t)
 \, 
 \label{rels}
\end{eqnarray}
%The mixing angles always have a characteristic hierarchy. For small mixing angles, and large VLQ mass $M$,
and 
\begin{eqnarray}
M_{T,B} \tan \theta_R^{t,b}&=m_{t,b}\tan \theta_L^{t,b}\qquad &{\hbox{singlets,~triplets}}\nonumber \\
M_{T,B} \tan \theta_L^{t,b}&=m_{t,b} \tan \theta_R^{t,b}\qquad &{\hbox{doublets}}\, .
\label{angrels}
\end{eqnarray}
Examples of the derivation of these relations are given in
Appendix~\ref{VLTapp} for the case of vector triplets. 

Except for the $(TB)$ doublet model, there are sufficient relationships that the results can always be expressed
in terms of two parameters.  For our numerical fits, we take as input parameters,
\begin{eqnarray}
&B~{\rm singlet:}&s_L^b, M_B\nonumber \\
&T~{\rm singlet:}&s_L^t, M_T\nonumber \\
&(XT)~{\rm doublet:}&s_R^t, M_T\nonumber \\
&(TB)~{\rm doublet:}&s_R^t,s_R^b, M_T\nonumber \\
&(BY)~{\rm doublet:}&s_R^b, M_B\nonumber \\
&(XTB)~{\rm triplet:}& s_L^t, M_T\nonumber \\
&(TBY)~{\rm triplet:}& s_L^t, M_T\, .
\label{pardef}
\end{eqnarray}

The couplings to the $W$ boson are,
\begin{eqnarray}
L_W&=&{g\over \sqrt{2}}\biggl({\overline q}_L^{i} \gamma_\mu
 A_{ij}^Lq_L^{j}+{\overline q} ^{i}_R\gamma_\mu A_{ij}^Rq_R^{j}\biggr)W^+_\mu + h.c.
\label{wcoups}
\end{eqnarray}
where $q^{i}, q^{j}$ are any two quarks in the model for which
$Q(q^i)-Q(q^j)=1$. 
%, and
%\begin{eqnarray}
%A_{ij}^L&=& (V_L^{u\dagger }V_L^d)_{ij}\nonumber \\
%A_{ij}^R&=& (V_R^{u\dagger }V_R^d)_{ij}\, .
%\end{eqnarray}
The values of $A_{ij}^{L,R}$ in the VLQ models we consider are 
reported in Tabs.~\ref{tab:ALVLQs} and~\ref{tab:ARVLQs}.

\begin{table}
\begin{center}
\begin{tabular}{|c|c|c|c|c|c|c|c|c|}
\hline\hline
 Model & $A^L_{tb}$ & $A^L_{tB}$ & $A^L_{TB}$ & $A^L_{Tb}$ & $A^L_{XT}$ & $A^L_{Xt}$ & $A^L_{BY}$ & $A^L_{bY}$
\\
\hline
 $T_s$ &$c_L^t$& & & $s_L^t$& & & &  \\
 \hline
 $B_s$ &$c_L^b$ & $s_L^b$ & &&&&&\\
 \hline
 $\psi_{XT}$&$c_L^t$& & & $s_L^t$&$c_L^t$& $-s_L^t$ & &\\
 \hline
 $\psi_{TB}$& $c_L^tc_L^b+s_L^ts_L^b$&$c_L^ts_L^b-s_L^tc_L^b$&$c_L^tc_L^b+s_L^ts_L^b$& $s_L^tc_L^b-c_L^ts_L^b$&&&&\\
 \hline
$\psi_{BY}$& $c_L^b$& $s_L^b$&&&&&$c_L^b$&$-s_L^b$\\
\hline
$\psi_{XTB}$& $c_L^tc_L^b+\sqrt{2}s_L^ts_L^b$ & $c_L^ts_L^b-\sqrt{2}s_L^tc_L^b$
&
$s_L^ts_L^b+\sqrt{2}c_L^tc_L^b$& $s_L^t c_L^b-\sqrt{2}c_L^ts_L^b$ & $\sqrt{2}c_L^t$& $-\sqrt{2}s_L^t$& &\\
\hline 
$\psi_{TBY}$&$c_L^tc_L^b+\sqrt{2}s_L^ts_L^b$ & $c_L^ts_L^b-\sqrt{2}s_L^tc_L^b$
&
$s_L^ts_L^b+\sqrt{2}c_L^tc_L^b$& $s_L^t c_L^b-\sqrt{2}c_L^ts_L^b$ & & &$\sqrt{2}c_L^b$&$-\sqrt{2}s_L^b$\\
\hline\hline
\end{tabular} 
 \caption{Left-handed fermion\,--\,$W$ couplings as defined in Eq.~(\ref{wcoups}).
 We assume all couplings are real, and neglect the SM CKM angles.} 
\label{tab:ALVLQs}
 \end{center}
 \end{table}

\begin{table}
\begin{center}
\begin{tabular}{|c|c|c|c|c|c|c|c|c|}
\hline\hline
 Model & $A^R_{tb}$ & $A^R_{tB}$ & $A^R_{TB}$ & $A^R_{Tb}$ & $A^R_{XT}$ & $A^R_{Xt}$ & $A^R_{BY}$ & $A^R_{bY}$
\\
\hline
 $T_s$ && & & $$& & & &  \\
 \hline
 $B_s$ & &  & &&&&&\\
 \hline
 $\psi_{XT}$&& & &&$c_R^t$& $-s_L^t$ & &\\
 \hline
 $\psi_{TB}$& $s_R^ts_R^b$&$-s_R^tc_R^b$&$c_R^tc_R^b$& $-c_R^ts_R^b$&&&&\\
 \hline
$\psi_{BY}$& & &&&&&$c_R^b$&$-s_R^b$\\
\hline
$\psi_{XTB}$& $\sqrt{2}s_R^ts_R^b$ & $-\sqrt{2}s_R^tc_R^b$
&
$\sqrt{2}c_R^tc_R^b$& $-\sqrt{2}c_R^ts_R^b$ & $\sqrt{2}c_R^t$& $-\sqrt{2}s_R^t$& &\\
\hline 
$\psi_{TBY}$&$\sqrt{2}s_R^ts_R^b$ & $-\sqrt{2}s_R^tc_R^b$
&
$\sqrt{2}c_R^tc_R^b$& $-\sqrt{2}c_R^ts_R^b$ & & &$\sqrt{2}c_R^b$&$-\sqrt{2}s_R^b$\\
\hline\hline
\end{tabular} 
 \caption{Right-handed fermion\,--\,$W$ couplings as defined in Eq.~(\ref{wcoups}).
 We assume all couplings are real.} 
\label{tab:ARVLQs}
\end{center}
\end{table}

The neutral current couplings to the 
$Z$ boson are also modified. The couplings for \mbox{$f_{i,j}=t,b,T,B,X,Y$} are, 
\begin{eqnarray}
{\cal L}_{Z}&=&{g\over 2 c_W}Z_\mu {\overline f}_i\gamma^\mu
\biggl[X_{ij}^L
P_L
+X_{ij}^R P_R - 2 Q_i \delta_{ij} s_W^2\biggr] f_j \, ,
\label{zdef}
\end{eqnarray}
where $s_W=\sin\theta_W$ is the weak mixing angle. 
% For $f=t_L (b_L)$, 
The SM couplings are normalized such that 
$X_{ij}^L=\delta_{ij}$ for $i=t$ and $X_{ij}^L=-\delta_{ij}$ for $i=b$ , with all other
$X$ equal to $0$.
%We note that $X^L_{uu}=(A^LA^{L\dagger})_{uu}$ and
%$X^L_{dd}=(A^{L\dagger} A^L)_{dd}$. 
%{\bf We note that~\cite{Lavoura:1992np}~$X^{L,R}_{ij}=A^{L,R}(A^{L,R})^\dagger$ 
%in the top sector and
%$X^{L,R}_{ij}=(A^{L,R})^\dagger  A^{L,R}$ in the bottom sector.}
For multiplets containing a heavy charge $-{1\over 3}$ quark with isospin $I_3^B$ that mixes 
with the SM-like $b$ quark or a heavy charge ${2\over 3}$ quark with isospin $I_3^T$ that mixes
with the SM-like $t$ quark, the diagonal  fermion couplings to the $Z$ are\footnote{$I_3=(2,0,-2)$ for triplets,
$(1,-1)$ for doublets, $0$ for singlets.},
\begin{eqnarray}
X_{ii}^{L}&=I_3^i(1-\delta_{iT})(1-\delta_{iB})+\delta X_{ii}^L\qquad &
X_{ii}^R= \delta X_{ii}^R\, , 
\end{eqnarray}
where the $I^3_i$ term in the left-handed couplings survive only for the
top and bottom quarks, and 
\begin{equation}
\label{xbbdef}
\begin{array}{lcllcl}
\delta X_{bb}^L&=& (s_L^b)^{2}(I_3^B+1)\qquad &\qquad
\delta X_{bb}^R&=&(s_R^b)^{2} I_3^B \\
\delta X_{tt}^L&=& (s_L^t)^{2}(I_3^T-1)\qquad &\qquad
\delta X_{tt}^R&=& (s_R^t)^{2} I_3^T \\
\delta X_{BB}^L&=& -1+(c_L^b)^{2}(I_3^B+1)\qquad& \qquad
\delta X_{BB}^R&=&(c_R^b)^{2} I_3^B \\
\delta X_{TT}^L&=& 1+(c_L^t)^{2}(I_3^T-1)\qquad &\qquad
\delta X_{TT}^R&=& (c_R^t)^{2} I_3^T \ \\
\delta X_{XX}^L&=&\delta X_{XX}^R=I_3^X \qquad&\qquad
\delta X_{YY}^L&=&\delta X_{YY}^R=I_3^Y\, .
\end{array}
\end{equation}
The off-diagonal couplings to the $Z$ boson are,
\begin{eqnarray}
X_{ij}^{L}&=\delta X_{ij}^L(1-\delta_{ij})\qquad &
X_{ij}^R= \delta X_{ij}^R(1-\delta_{ij})\, , 
\end{eqnarray}
where
\begin{eqnarray}
\delta X_{bB}^L&= -s_L^b c_L^b(I_3^B+1)\qquad &
\delta X_{bB}^R=-s_R^b c_R^b I_3^B\nonumber \\
\delta X_{tT}^L&= -s_L^t c_L ^t(I_3^T-1)\qquad & 
\delta X_{tT}^R= - s_R^t c_R^t I_3^T\, . 
\end{eqnarray}
Finally, the couplings to the Higgs boson can be parameterized as,
\begin{equation}
L=-{h\over v} {\overline f}_{L}^i c_{ij}f_{R}^j+h.c.  \;.
\label{higgscoup}
\end{equation}
The flavor non-diagonal fermion-Higgs couplings are important for double Higgs
production~\cite{Dawson:2012mk,Gillioz:2012se} and can be found in Ref.~\cite{Aguilar-Saavedra:2013qpa}.
For models with a singlet or triplet VLQ,
\begin{equation}
c_{ij}=V_L {\cal{F}} V_L^\dagger M_{diag}
\end{equation}
and for models with a  doublet VLQ,
\begin{equation}
c_{ij}=M_{diag} V_R {\cal{F}} V_R^\dagger ,
\end{equation}
where
\begin{equation}
{{\cal F}}\equiv\left(\begin{matrix}
1&0\\0&0\end{matrix}\right)\, .
\end{equation}
These formulae hold for both the charge ${2\over 3}$ and charge $-{1\over 3}$
sectors. The $X$ and $Y$ fermions do not couple to the Higgs. 
The diagonal Higgs couplings are given in Table \ref{tab:hcoups}. 
\begin{table}
\begin{center}
\begin{tabular}{|c|c|c|c|c|}
\hline\hline
 Model & $c_{bb}$& $c_{BB}$ & $c_{tt}$ & $c_{TT}$ \\
 \hline\hline
 $T_s$ &$m_b$&-&$m_t(c_L^t)^{2}$&$M_T(s_L^t)^{2}$\\
 \hline
 $B_s$ &$m_b(c_L^b)^{2}$&$M_B(s_L^b)^{2}$&$m_t$&-\\
 \hline
 $\psi_{XT}$&$m_b$&-&$m_t(c_R^t)^{2}$&$M_T(s_R^t)^{2}$\\
 \hline
 $\psi_{TB}$&$m_b(c_R^b)^{2}$&$M_B(s_R^b)^{2}$&$m_t(c_R^t)^{2}$&$M_T(s_R^t)^{2}$\\
\hline
$\psi_{BY}$&$m_b(c_R^b)^{2}$&$M_B (s_R^b)^{2}$&$m_t$&-\\
\hline
$\psi_{XTB}$&$m_b(c_L^b)^{2}$&$M_B(s_L^b)^{2}$&$m_t(c_L^t)^{2}$&$M_T(s_L^t)^{2}$\\
\hline 
$\psi_{TBY}$&$m_b(c_L^b)^{2}$&$M_B(s_L^b)^{2}$&$m_t(c_L^t)^{2}$&$M_T(s_L^t)^{2}$\\
\hline\hline
\end{tabular} 
 \caption{\label{tab:hcoups} Diagonal Higgs couplings to fermions.} 
 \end{center}
 \end{table}
 
\section{VLQ Contributions to Precision Measurements}
\label{vlqob}
Electroweak precision data place strong restrictions on the parameters of models with VLQs. In this section,
we review the contributions to the oblique parameters and the $Z b {\overline b}$ couplings in the VLQ models
introduced in the previous section. 

\subsection{Oblique Parameters}
The general expression for the contribution to the  $T$ parameter from fermions 
is~\cite{Lavoura:1992np,Anastasiou:2009rv,Chen:2003fm}
\begin{eqnarray}
T&=& {N_c\over 16 \pi s_W^2 c_W^2}\sum_{i.j}\biggl\{ 
\biggl(\mid A_{ij}^L\mid^2+\mid A_{ij}^R\mid^2\biggr)\theta_+(y_i, y_j)
+2 {\rm Re}\biggl(A_{ij}^L
A_{ij}^{R*}\biggr)\theta_- (y_i,y_j)
\nonumber \\
&&-{1\over 2}\biggr[\biggl(\mid X_{ij}^L\mid^2
+\mid X_{ij}^R\mid^2\biggr)\theta_+(y_i, y_j)
+2 {\rm Re}\biggl(X_{ij}^L
X_{ij}^{R*}\biggr)\theta_- (y_i,y_j)\biggr]\biggr\}\, ,
\label{eq:dtdef}
\end{eqnarray}
where  $N_c=3$,  $y_i\equiv {M_{Fi}^2\over M_Z^2}$, $M_{F_i}$ are the fermion masses, and  $A^{L,R}_{ij}$, $X^{L,R}_{ij}$
are defined in Eqs.~\ref{wcoups} and~\ref{zdef} respectively. For the
input parameters we use~\cite{Olive:2016xmw} $m_t = 173.5$~GeV,
$m_b=4.2$~GeV, $m_Z =91.1876~{\rm GeV}, m_W = 80.385~{\rm GeV}$ and define the weak angle
through $ c_W ={ m_W \over m_Z}$.\\
The functions $\theta_\pm (y_i,y_j)$ are,
\begin{eqnarray}
\theta_+(y_1,y_2)&=& y_1+y_2-{2y_1y_2\over y_1-y_2}\log\biggl({y_1\over y_2}\biggr)\\
\theta_-(y_1,y_2)&=& 2\sqrt{y_1y_2}
\left[{y_1+y_2\over y_1 -y_2}\ln\biggl({y_1\over y_2}\biggr)-2\right]\, .
\end{eqnarray}
We note that $\theta_+(y,y)=\theta_-(y,y)=0$. When  $y_1>> y_2$, $\theta_+(y_1,y_2)\xrightarrow[]{y_1>>y_2} y_1$
and $\theta_-(y_1,y_2) \xrightarrow[]{y_1>>y_2}  0$. We will make use
of these properties as we compute all electroweak parameters in the limit $m_b<< m_t$.

As
customary, we subtract the SM top-bottom contribution,
\begin{eqnarray}
\Delta T&=&T-T_{SM}
\end{eqnarray}
where
\bea
T_{SM}&=& {N_c\over 16 \pi s_W^2 c_W^2} \theta_+(y_t,y_b)%\biggl\{ \theta_+(y_t,y_b)
%-{1\over 2}\biggl[ \theta_+(y_t,y_t)+\theta_+(y_b,y_b)\biggr]\biggr\}
=  {N_c\over 16 \pi s_W^2 }{m_t^2 \over m_W^2} 
\, .
\end{eqnarray}
\\
For the top and bottom singlet partner models, the exact results are simple~\cite{Lavoura:1992np,Dawson:2012di}
\begin{eqnarray}
T~{\rm singlet:} & \Delta T^{T_s}=&{N_c m_t^2 \over 16 \pi s_W^2 M_W^2} (s_L^t)^2
	\left[ - \left(1+(c_L^t)^2\right) +2 (c_L^t)^2  {r_T\over r_T-1} \log(r_T)  + (s_L^t)^2 r_T\right]
	\nonumber \\ 
	& &\\ 
B~{\rm singlet:} & \Delta T^{B_s}=&{N_c m_t^2\over 16 \pi s_W^2 M_W^2}  (s_L^b)^2 r_B 
\left[{2\over 1-r_B} \log(r_B) + (s_L^b)^2 \right] \;,
\end{eqnarray}
where $r_F\equiv{M_F^2\over m_t^2}$.
The contribution  to the $T$ parameter in the $(BY)$ doublet model also has a simple expression:
\begin{eqnarray}
(BY)~{\rm doublet:}~\Delta T^{BY}&=&
-{N_c m_t^2\over 128\pi s_W^2 M_W^2} r_B
\biggl\{  32 {(c^b_R)^2 \over
    s^b_R} \log (c_R^b) \left[
    (c^b_R)^2+1 \right] \nonumber \\
&& +8 s_R^b \left[4 (c^b_R)^2 - (s_R^b)^4\right]\biggr\}
\,   .
\end{eqnarray}

In the large  VLQ mass  and small mixing angle limits, we obtain simple approximate 
expressions for the $T$ parameter for all the VLQ representations\footnote{ Exact results are 
posted  at \href{$quark.phy.bnl.gov\backslash Digital\_Data\_Archive\backslash dawson\backslash vlq\_17$}{https://quark.phy.bnl.gov/Digital_Data_Archive/dawson/vlq_17/}
as a mathematica notebook.}$^,$\footnote{In all our studies we will use
the exact expressions for the Peskin-Takeuchi parameters, retaining
the full dependence on the VLQs masses and on the mixing angles.}
\begin{eqnarray}
T~{\rm singlet:} & \Delta T^{T_s}\sim & {N_c m_t^2 \over 8 \pi s_W^2 M_W^2}
	(s_L^t)^2 \left[ 	\log(r_T) - 1 \right]+{\cal O}\biggl((s_L^t)^4,{1\over r_T}\biggr)
	 \nonumber \\
B~{\rm singlet:} & \Delta T^{B_s}\sim &- {N_c m_t^2 \over 8 \pi s_W^2 M_W^2}
	(s_L^b)^2 	\log(r_B)+{\cal O}\biggl((s_L^b)^4,{1\over r_B}\biggr)  \nonumber \\
(TB)~{\rm doublet:} & \Delta  T^{TB} \sim & {N_c m_t^2(s_R^t)^2\over 8 \pi s_W^2 M_W^2}
\biggl[
-3+2\log(r_T) \biggr] +\nonumber \\ 
&&{\cal O}\biggl((s_R^t)^4,  (s_R^t)^2  (s_R^b)^2 , (s_R^b)^4,
                                                    {1\over
   r_T}\biggr)  \nonumber \\
(XT)~{\rm doublet:} & \Delta  T^{XT} \sim& {N_c m_t^2(s_R^t)^2
\over 8\pi s_W^2 M_W^2}
\biggl[3-2\log(r_T)\biggr]+
{\cal O}\biggl((s_R^t)^4,{1\over r_T}\biggr)
\nonumber \\
(BY)~{\rm doublet:} & \Delta T^{BY} \sim& 
{N_c  m_t^2( s_R^b)^5 r_B\over 12\pi s_W^2 M_W^2}
+ {\cal O}\biggl((s_R^b)^7,{1\over r_B}\biggr)\nonumber 
\eea
\bea
(XTB)~{\rm triplet:}&\Delta T^{XTB}\sim&{N_c m_t^2\over 8 \pi s_W^2 M_W^2}(s_L^t)^2\biggl[3\log(r_T)-5\biggr]
+{\cal O}\biggl((s_L^t)^4,{1\over r_T}\biggr)\nonumber \\
(TBY)~{\rm triplet:}&\Delta T^{TBY}\sim &-{N_c m_t^2\over 16 \pi s_W^2 M_W^2}3(s_L^t)^2\biggl[\log(r_T)-2\biggr] 
%\nonumber\\
+{\cal O}\biggl((s_L^t)^4,{1\over r_T}\biggr)\; .\nonumber\\
\label{tdef_approx}
\end{eqnarray}

\begin{figure}[tb]
\centering
\includegraphics[width=0.45\textwidth,clip]{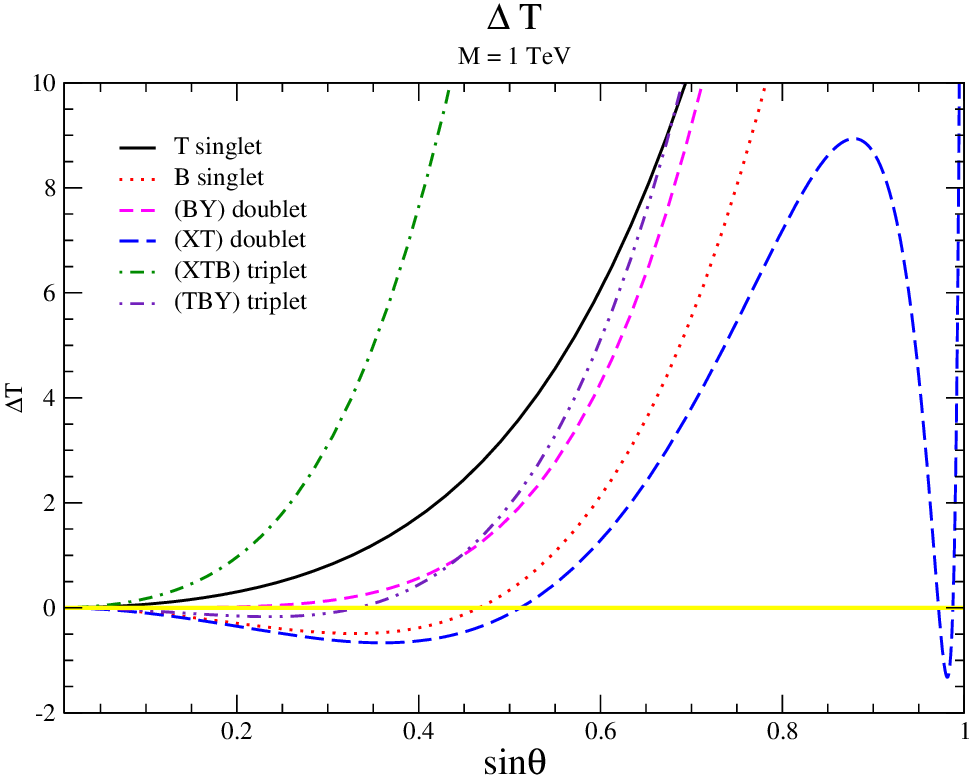}
\includegraphics[width=0.45\textwidth,clip]{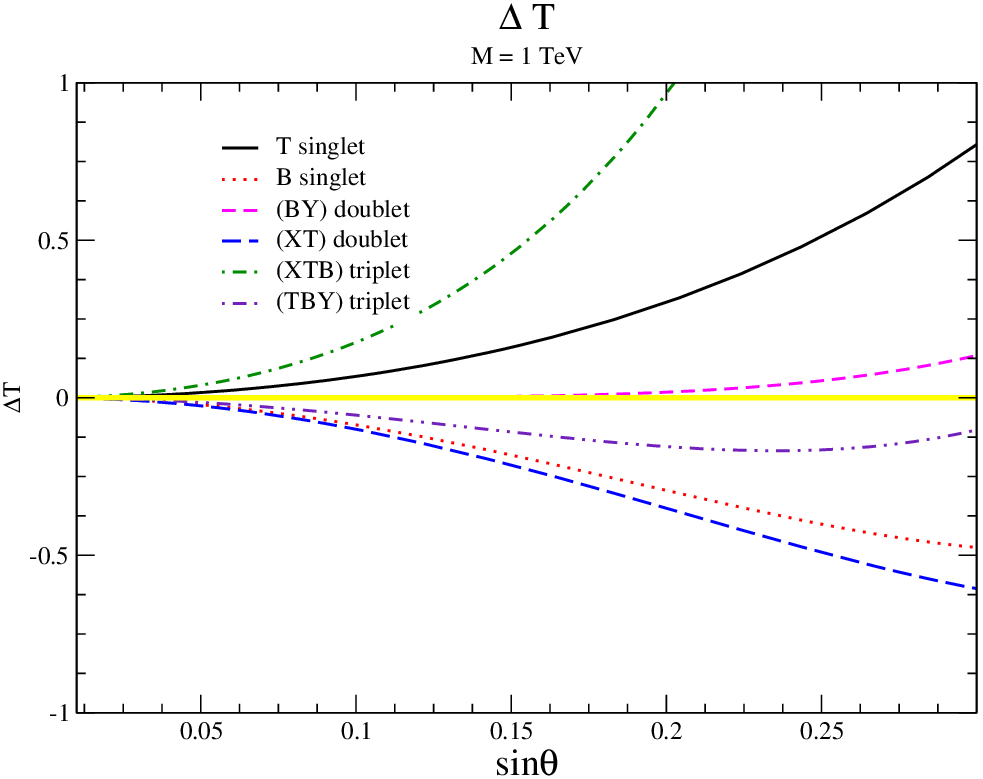}
\caption{Exact results for $\Delta T$ for $M=1~$TeV in the VLQ models. $\sin\theta$ and $M$
are identified in Eq.~\ref{pardef}.}
\label{fig:dt}
\end{figure}

\begin{figure}[tb]
\centering
\includegraphics[width=0.45\textwidth,clip]{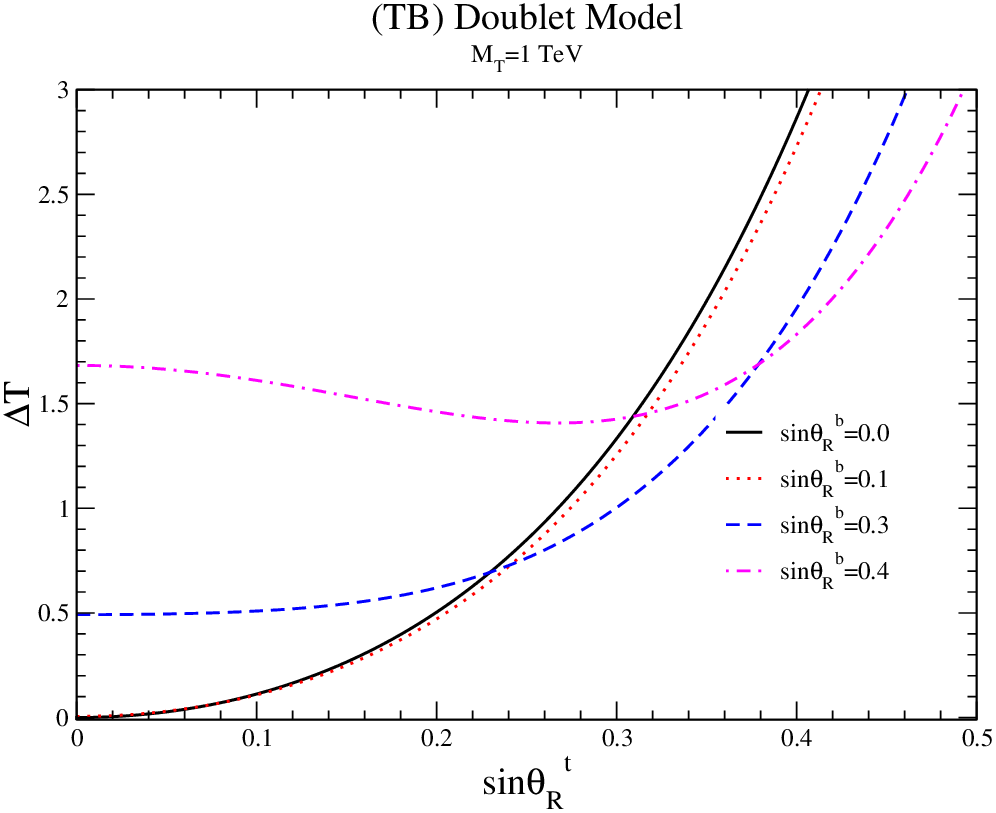}
\caption{Exact results for $\Delta T$ for $M_T=1~$TeV in the $(TB)$ doublet model.}
\label{fig:dt_tbmod}
\end{figure}

The contributions to $\Delta T$ in the various VLQ models %with a single multiplet
are shown in Fig.~\ref{fig:dt} (Fig.~\ref{fig:dt_tbmod} for the $(TB)$
doublet, which has two mixing angles as free parameters). Here we use the exact expressions for
the $T$ parameter. % Eq. \ref{eq:dtdef}. 
For small mixing, the contribution to $\Delta T$ is positive in the $T$ singlet and $(XTB)$ triplet models,
negative in the $B$ singlet, $(XT)$ doublet and $(TBY)$ triplet models, and extremely small in the $(BY)$ 
doublet model (RHS of Fig.~\ref{fig:dt}), as one could expect from the
approximate results in Eq.~\ref{tdef_approx}. 
In all the models where the $T$ parameter is negative for small mixing 
%have the interesting feature that 
$\Delta T$ changes sign at an
intermediate value of $\sin\theta$ and therefore vanishes again for non-small
mixing. In the case of the $(XT)$ doublet,  $\Delta T^{XT} \sim 0$ even 
for $s_R^t \sim 1$, due to a numerical cancellation. 
Therefore, in these models there could be regions of parameter space with 
quite sizeable mixing that are allowed by precision tests. 
We will explore this possibility in Sec. IV.

The mass splitting between the VLQ multiplet
components, %$\delta\equiv M_T-M_B, ~M_X-M_T,~M_B-M_Y$, 
$\delta_{Q_1Q_2}\equiv M_{Q_1}-M_{Q_2}$, is fixed by the 
 mixing angles  (Eq.~\ref{rels}).  In the large VLQ mass and small
 mixing angle  approximation,  and in the limit for massless bottom
 quark, 
\begin{eqnarray}
{\rm Doublets:}\quad &
{\delta_{TB}\over M_T}&\sim  {1\over 2 }
\biggl[(s_R^t)^2\biggl(1-{m_t^2\over M_T^2}\biggr)-(s_R^b)^2\biggr]
\nonumber \\
\quad &{\delta_{XT}\over M_T} & \sim-{(s_R^t)^2\over 2}
\biggl(1-{m_t^2\over M_T^2}\biggr) < 0\nonumber \\
\quad & {\delta_{BY}\over M_B}&\sim {(s_R^t)^2\over 2} < {1\over 2}\nonumber \\
\end{eqnarray}
\begin{eqnarray}
{\rm Triplets:}\quad & {\delta_{TB}\over M_T}&\sim  {1\over 2 }
\biggl[(s_L^t)^2\biggl(1-{m_t^2\over M_T^2}\biggr)-(s_L^b)^2\biggr]\nonumber \\
\quad &{\delta_{XT}\over M_T} & \sim-{(s_L^t)^2\over 2}
\biggl(1-{m_t^2\over M_T^2}\biggr) < 0\nonumber \\
\quad & {\delta_{BY}\over M_B}&\sim {(s_L^b)^2\over 2} < {1\over 2}
\, .
\label{sigrels}
\end{eqnarray}
From Eq.~\ref{sigrels}, in all cases %, the mass splitting 
${\delta\over M}\sim \sin^2 \theta_i$, so for small angles $\Delta T$ grows with the mixing between the SM fermions
and the VLQs. For large masses, the mixing goes to zero for fixed
Yukawa couplings (see Table~\ref{tab:hcoups}), and decoupling is recovered~\cite{Dawson:2012di}.

The expression for the contributions to the $S$ parameter from fermions 
is~\cite{Lavoura:1992np,Carena:2007ua,Chen:2003fm}
%\footnote{This is in disagreement with the result of Ref.~\cite{AguilarSaavedra:2002kr}.},
\begin{eqnarray}
%S&=& {N_c\over 2 \pi}\Sigma_{\alpha,\beta}\biggl\{ 
%\biggl(\mid A_{\alpha\beta}^L\mid^2+\mid A_{\alpha\beta}^R\mid^2\biggr)\psi_+(y_\alpha, y_\beta)
%+2 Re\biggl(A_{\alpha\beta}^L
%A_{\alpha\beta}^{R*}\biggr)\psi_- (y_\alpha,y_\beta)
%\nonumber \\
%&&-{1\over 2}\biggr[\biggl(\mid X_{\alpha\beta}^L\mid^2
%+\mid X_{\alpha\beta}^R\mid^2\biggr)\chi_+(y_\alpha, y_\beta)
%+2 Re\biggl(X_{\alpha\beta}^L
%X_{\alpha\beta}^{R*}\biggr)\chi_- (y_\alpha,y_\beta)\biggr]\biggr\}\, ,
%
S&=& {N_c\over 2 \pi}\sum_{i,j}\biggl\{ 
\biggl(\mid A_{ij}^L\mid^2+\mid A_{ij}^R\mid^2\biggr)\psi_+(y_i, y_j)
+2 {\rm Re}\biggl(A_{ij}^L
A_{ij}^{R*}\biggr)\psi_- (y_i,y_j)
\nonumber \\
&&-{1\over 2}\biggr[\biggl(\mid X_{ij}^L\mid^2
+\mid X_{ij}^R\mid^2\biggr)\chi_+(y_i, y_j)
+2 {\rm Re}\biggl(X_{ij}^L
X_{ij}^{R*}\biggr)\chi_- (y_i,y_j)\biggr]\biggr\}\, ,
\end{eqnarray}
where we subtract the SM top-bottom contribution,
\begin{eqnarray}
\Delta S&=&S-S_{SM}\\
S_{SM}&=& {N_c\over 6 \pi}\biggl[1-{1\over 3}\log\biggl({m_t^2\over m_b^2}\biggr)\biggr]\, .
\end{eqnarray}
The functions appearing in $S$ are~\cite{Lavoura:1992np},
\bea
   \psi_+(y_1,y_2)&=&{1\over 3} -{1\over 9} \log{y_1 \over y_2}
     \nonumber \\
   \psi_-(y_1,y_2)&=&-{y_1+y_2 \over 6 \sqrt{y_1 y_2}} 
     \nonumber \\
   \chi_+(y_1,y_2)&=&{5(y_1^2+y_2^2)-22 y_1 y_2\over 9 (y_1-y_2)^2}
      +{3 y_1 y_2  (y_1+y_2)- y_1^3-y_2^3\over 3 (y_1-y_2)^3} \log
      {y_1 \over y_2} 
      \nonumber \\
  \chi_-(y_1,y_2)&=& -\sqrt{y_1 y_2} \left[ {y_1+y_2\over 6 y_1 y_2}-
      {y_1+y_2 \over  (y_1-y_2)^2} +{2 y_1 y_2 \over (y_1-y_2)^3} \log
      {y_1 \over y_2}\right]
\eea
% OLD/typos
%\begin{eqnarray}
%\psi_+(y_1,y_2)&=&{22y_1+14y_2\over 9}-{1\over 9}
%\log\biggl({y_1\over y_2}\biggr)+{11y_1+1\over 18}f(y_1,y_1)
%+{7y_2-1\over 18}f(y_2,y_2)\nonumber\\
%\chi_+(y_1,y_2)&=&{y_1+y_2\over 2}
%-{(y_1-y_2)^2\over 3}+\biggl[{(y_1-y_2)^3\over 6}
%-{1\over 2}{y_1^2+y_2^2\over y_1-y_2}\biggr]\log\biggl({y_1\over y_2}\biggr)
%\nonumber \\
%&&+{y_1-1\over 6}f(y_1,y_1)+{y_2-1\over 6}f(y_2,y_2)\nonumber \\
%\psi_-(y_1,y_2)&=& -\sqrt{y_1 y_2} \biggl[ 4+{1\over 2}\biggl(f(y_1,y_1)+f(y_2,y_2)\biggr)
%\biggr]
%\nonumber \\ 
%\chi_-(y_1,y_2)&=& -\sqrt{y_1,y_2}\biggl[2+\biggl(y_1-y_2-{y_1+y_2\over y_1-y_2}\biggr)\ln\biggl({y_1\over y_2}\biggr)
%\nonumber \\ 
%&&+{1\over 2} \biggl(f(y_1,y_1)+f(y_2,y_2)-f(y_1,y_2)\biggr)\biggr]\nonumber \\
%f(y_1,y_2)&=&-2Re\int_0^1dx\biggl\{\log\biggl(y_1 x+y_2(1-x)-x(1-x)\biggr)
%\nonumber \\ &&
%-4 +2(1+y_1-y_2)\ln\biggl(
%{y_1\over y_2}\biggr)\biggr\}\, , 
%\end{eqnarray}
where 
$ \chi_+(y,y)=\chi_-(y,y) = 0$ and 
in the limit $y_1>>y_2$,
\begin{eqnarray}
\psi_+(y_1,y_2)&\xrightarrow[]{y_1>>y_2}  
{1\over 3}-{1\over 9} \log\biggl(
{y_1\over y_2}
\biggr), & \psi_-(y_1,y_2) \xrightarrow[]{y_1>>y_2} 
-{1\over 6}\sqrt{{y_1\over y_2}}
\nonumber \\
\chi_+(y_1,y_2)&\xrightarrow[]{y_1>>y_2}  {5\over 9}-
{1\over 3}
\log\biggl( {y_1\over y_2}
\biggr),
& \chi_-(y_1,y_2) \xrightarrow[]{y_1>>y_2} -{1\over 6}\sqrt{{y_1\over y_2}}
\end{eqnarray}

For the singlet bottom and top VLQ models, the exact results (full mass and angle dependence)
are
\begin{eqnarray}
T~{\rm singlet:} & \Delta S^{T_s}=&
 -{N_c\over 18 \pi} (s_L^t)^2
 \left[ \log(r_T)+(c_L^t)^2 \left(
 	{5(r_T^2+1)-22r_T\over (r_T-1)^2} \right.\right.\\ && 
	\left. \left. +{3(r_T+1)(r_T^2-4r_T+1)\over (1-r_T)^3}\log(r_T)\right)\right] \;,
\nonumber \\
B~{\rm singlet:} & \Delta S^{B_s}=& {N_c \over 18 \pi} (s_L^b)^2 \left[ -5 (c_L^b)^2
	+\left(4-3(s_L^b)^2\right) \log{r_B \over r_b}\right] \;.
\end{eqnarray}
As in the case of the $T$ parameter, the $(BY)$ doublet model also
has a simple exact expression for $S$,
\bea
    (BY)~{\rm doublet:} & \Delta S^{BY}=&
    {N_c\over 18 \pi} \left\{
         4 (c_R^b)^2 \log (c_R^b)+
         (s_R^b)^2  \left[ 1+ 5 (s_R^b)^2+\left(2-3 (s_R^b)^2 \right)
           \log {r_B\over r_b} \right] \right\} \;. \nonumber\\
\eea

\begin{figure}[tb]
\centering
\includegraphics[width=0.45\textwidth,clip]{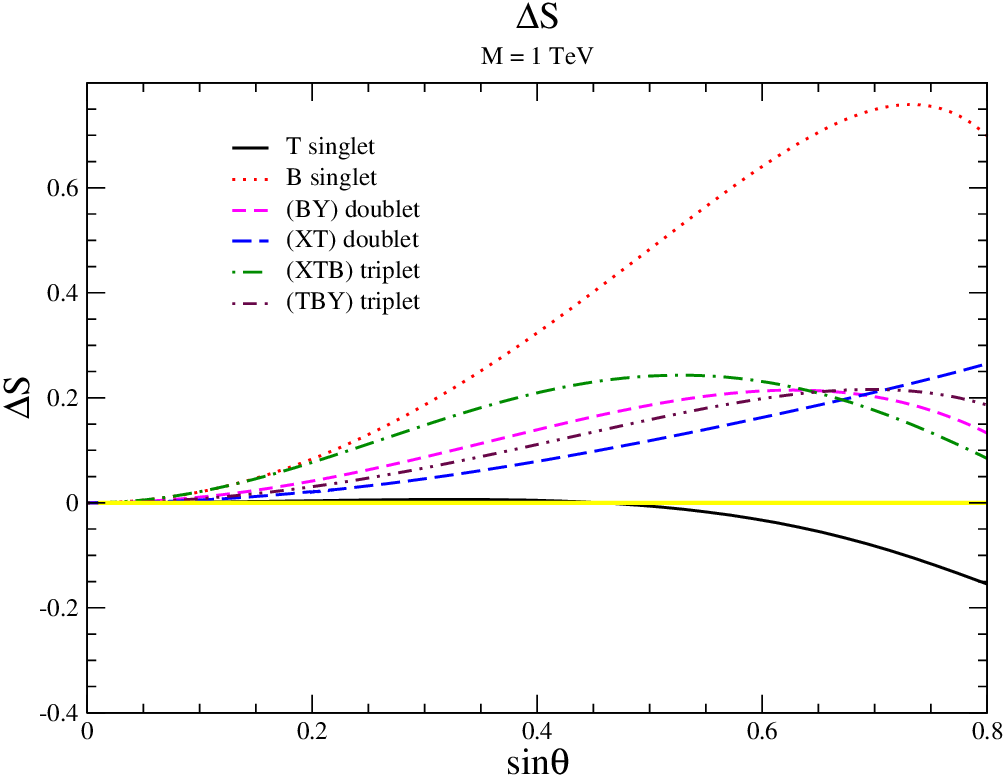}
\includegraphics[width=0.45\textwidth,clip]{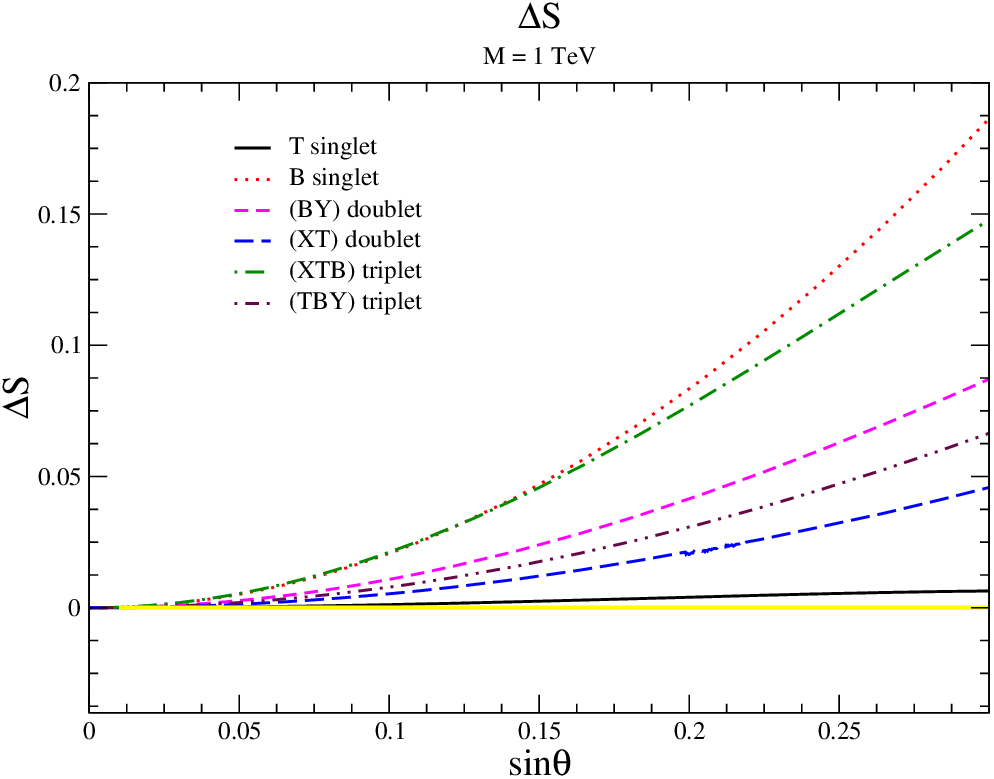}
\caption{Exact results for $\Delta S$ for $M=1~$TeV in the VLQ 
models. The parameters $\sin\theta$ and $M$
are identified in Eq.~\ref{pardef}.}
\label{fig:ds}
\end{figure}

\begin{figure}[tb]
\centering
\includegraphics[width=0.4\textwidth,clip]{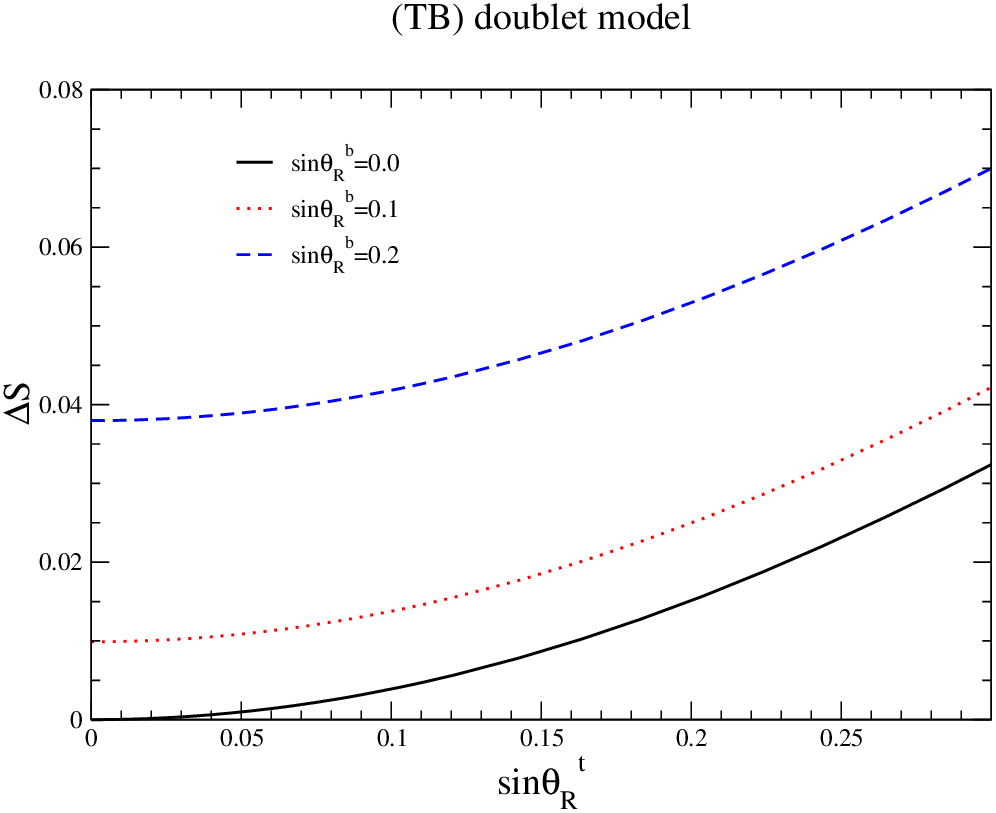}
\caption{Exact results for $\Delta S$ for $M=1~$TeV in the $(TB)$ doublet model.}
\label{fig:ds_tbmod}
\end{figure}

The contributions to $\Delta S$ are shown in Figs.~\ref{fig:ds}
and~\ref{fig:ds_tbmod} using the exact results (full mass and
angle dependence) in all the models.
%Eq.~\ref{eq:dtdef}. 


In the heavy VLQ mass limit (and assuming small mixings between
the doublet and triplet components),\footnote{ Exact results are 
posted  at \href{$quark.phy.bnl.gov\backslash Digital\_Data\_Archive\backslash dawson\backslash vlq\_17$}{https://quark.phy.bnl.gov/Digital_Data_Archive/dawson/vlq_17/}
as a mathematica notebook.}
\begin{eqnarray}
	T~{\rm singlet:} & \Delta S^{T_s}\sim & {N_c \over 18 \pi} (s_L^t)^2 
		\left[ -5 + 2 \log( r_T) \right] + {\cal O}\biggl((s_L^t)^4, {1\over r_T}\biggr)  
		\nonumber \\
	B~{\rm singlet:} & \Delta S^{B_s}\sim& {N_c \over 18 \pi} (s_L^b)^2 
		\left[ -5 + 4 \log{r_B \over r_{b}}\right]+{\cal O}\biggl((s_L^b)^4,{1\over r_B}\biggr) 
		\nonumber 
\eea
\bea
(TB)~{\rm doublet:} & \Delta  S^{TB} \sim&
                       - {N_c\over 18 \pi} 
                            \left\{  (s_R^b)^2 \left[3 + 2 \log\left( {r_b \over r_T}\right) \right]
                            +(s_R^t)^2 \left[7 - 4 \log(r_T)\right]
                            \right\} +\nonumber \\ 
                         &&{\cal O}\biggl((s_R^t)^4,  (s_R^t)^2  (s_R^b)^2 , (s_R^b)^4,
                                                    {1\over r_T}\biggr) 
                 \nonumber \\
(XT)~{\rm doublet:} & \Delta  S^{XT} \sim& {N_c\over 18\pi}(s_R^t)^2 \biggl[3+2\log(r_T) 
\biggr]+ {\cal O}\biggl((s_R^t)^4, {1\over r_T}\biggr)  
\nonumber \\
(BY)~{\rm doublet:} & \Delta  S^{BY} \sim&{N_c(s_R^b)^2\over 18\pi }
\biggl[-1+2\log\biggl({r_B\over r_{b}}\biggr)\biggr] +{\cal O}\biggl((s_R^b)^4,{1\over r_B}\biggr) 
\nonumber \\
(XTB)~{\rm triplet:} & \Delta S^{XTB} \sim & -{N_c\over 18\pi} (s_L^t)^2 \biggl[ 7+4\log(r_{b})-
6\log(r_T)\biggr] + {\cal O}\biggl((s_L^t)^4, {1\over r_T}\biggr)  
\nonumber \\
(TBY)~{\rm triplet:} & \Delta S^{TBY} \sim & {N_c\over 36\pi} (s_L^t)^2 \biggl[ 1+8\log(r_{T})\biggr] 
+ {\cal O}\biggl((s_L^t)^4, {1\over r_T}\biggr)  
\, .
\end{eqnarray}
%where $r_B={M_B^2\over m_t^2}$ and $r_{b}={m_b^2\over m_t^2}$.

\begin{figure}[tb]
\centering
\includegraphics[width=0.45\textwidth,clip]{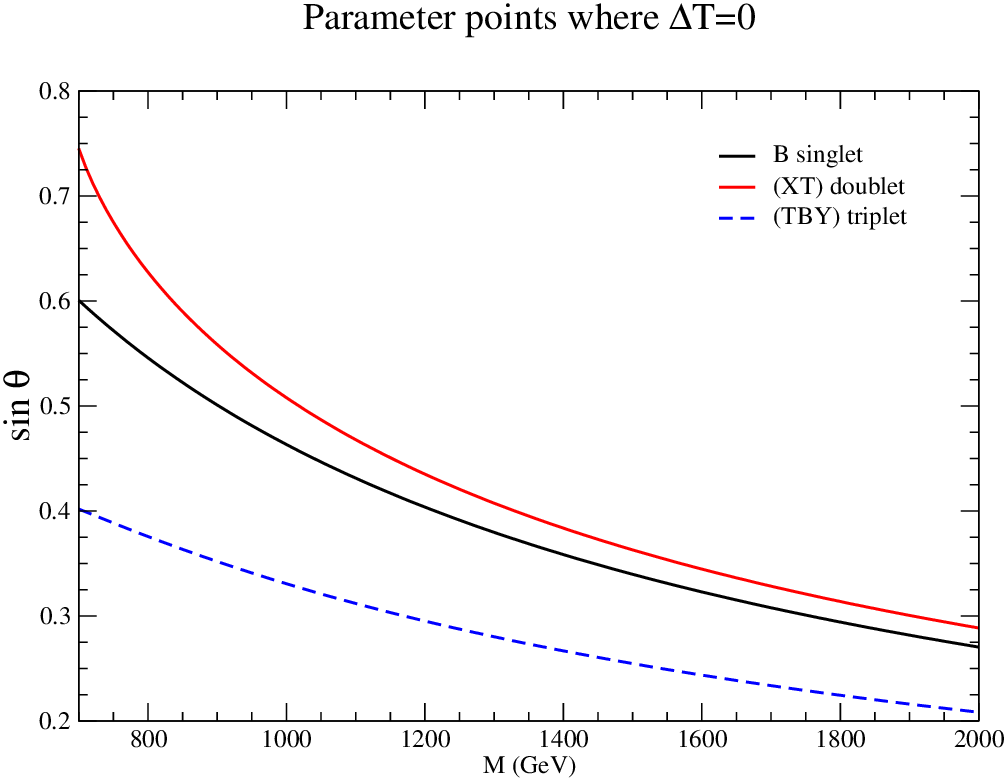}
\includegraphics[width=0.45\textwidth,clip]{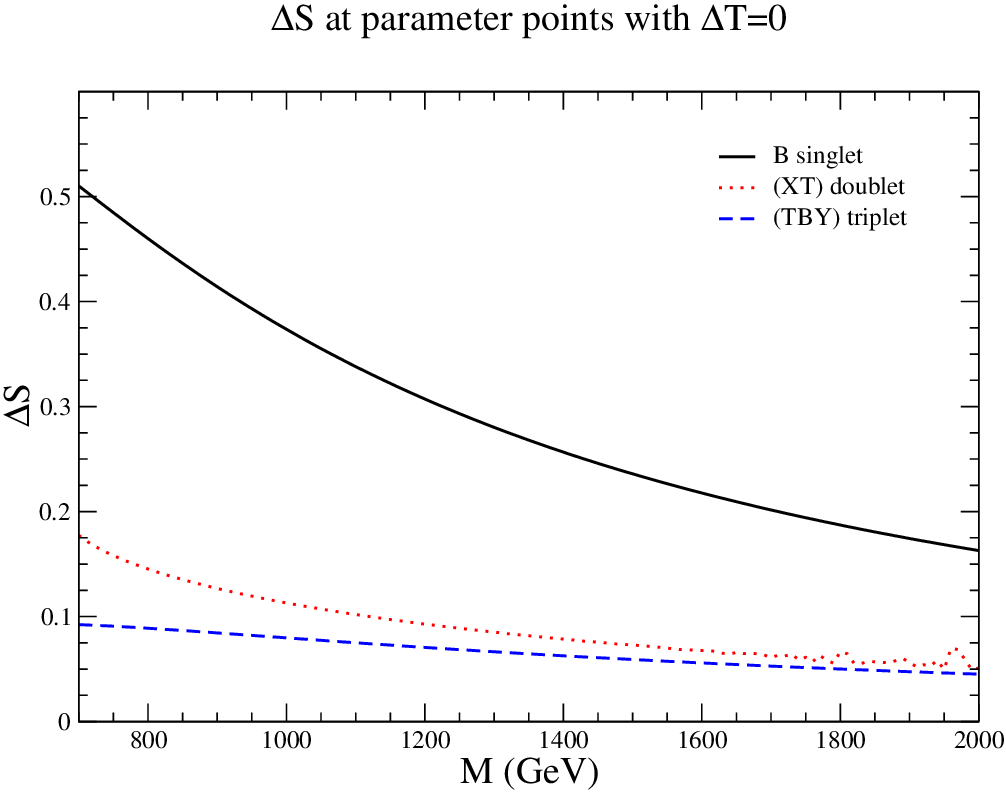}
\caption{Left panel: fine-tuned parameter points where $\Delta T=0$ in the $B$ singlet, 
$(XT)$ doublet and $(TBY)$ triplet models. The parameters $\sin\theta$ and $M$
are identified in Eq.~\ref{pardef}. Right panel: values of the $S$
parameter corresponding to the points on the LHS.}
\label{fig:bsing_t0}
\end{figure}

The $B$ singlet, $(XT)$ doublet, and $(TBY)$ triplet models have the interesting feature
that $\Delta T$ vanishes for particular  fine-tuned
choices of the parameters with non-zero mass splittings between the members of the VLQ multiplets.
In the left panel of Fig.~\ref{fig:bsing_t0} 
we show the VLQ mass and mixing angle for which $\Delta T=0$ and in
the right panel we show $\Delta S$
for these parameters. %in the $B$ singlet model, the $(XT)$ doublet and $(TBY)$ triplet models.   

\begin{figure}[tb]
\centering
\includegraphics[width=0.6\textwidth,clip]{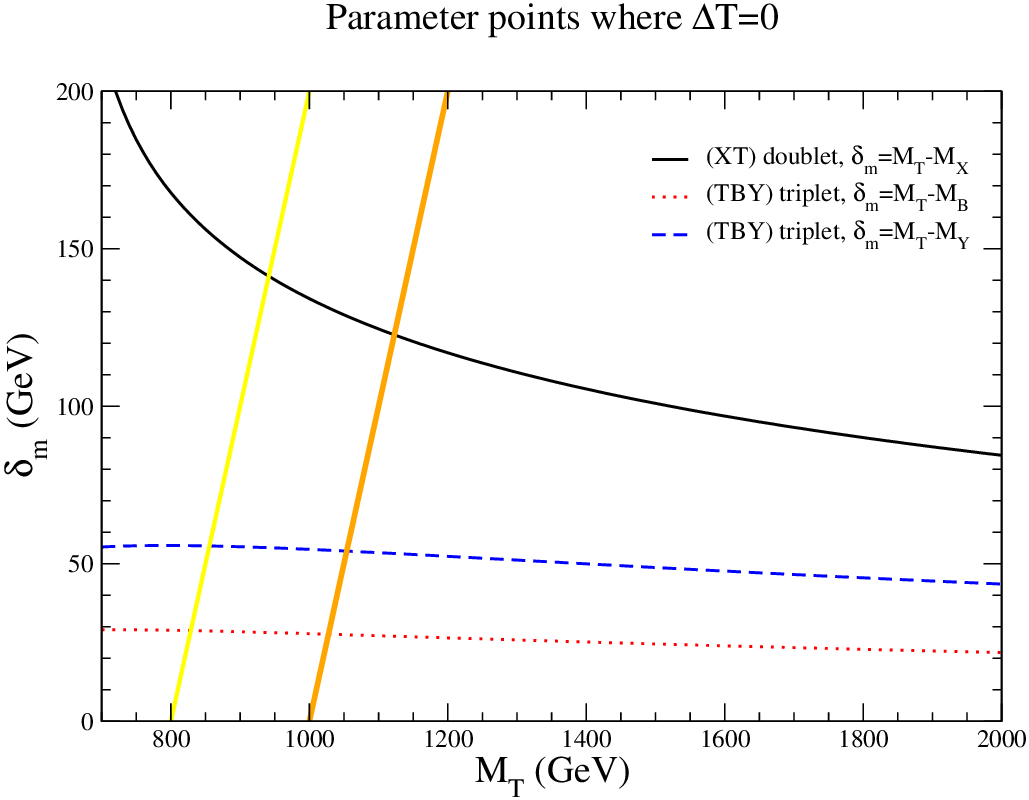}
\caption{VLQ multiplet mass splittings for parameter points where $\Delta T=0$.
Below and to the right of the yellow (orange)  line, all VLQs have masses larger than 800~GeV
(1~TeV).}
\label{fig:deltm_t0}
\end{figure}

The oblique fit, ignoring correlations,
requires $\Delta S < 0.3$ at $95~\%$ confidence level~\cite{deBlas:2016ojx,Baak:2014ora}, so there are
regions where both $\Delta T$ and $\Delta S$ can escape the oblique constraints in these models.  These
fined-tuned regions 
will have important impacts on the global fits in the next section and we note
that the mass splittings between VLQ multiplet members
can be significant for these choices of parameters.  
Fig.~\ref{fig:deltm_t0} shows the mass difference for
the points where $\Delta T$ is fine-tuned to be zero, corresponding to the mixing 
angles of Fig.~\ref{fig:bsing_t0}.

%\begin{figure}[tb]
%\centering
%\includegraphics[width=0.4\textwidth,clip]{dt_acc_tb.eps}
%\includegraphics[width=0.4\textwidth,clip]{ds_acc_tb.eps}
%\caption{Comparison between exact and approximate results in the $(TB)$ doublet model.}
%\label{fig:stu_tb}
%\end{figure}
\subsection{Contributions to $Zb {\overline{b}}$}
In the VLQ models where $\delta X_{bb}^L$ vanishes at tree level, 
we will use the one-loop contributions to the left-handed  
$Zb{\overline{b}}$ coupling for our fit to electroweak precision data.  
This occurs in the $(TB)$~doublet model (Eq.~\ref{xbbdef}), as well 
as in the $T$~singlet and $(XT)$~doublet models. The one-loop corrections from
$t-T$ mixing 
to $\delta X_{bb}^L$ are~\cite{Burgess:1993vc,Anastasiou:2009rv},
\begin{equation}
\delta X_{bb}^L=
{g^2\over 32 \pi^2} 
(s_L)^2\biggl(
f_1(x,x^\prime)+(c_L^t)^2
f_2(x,x^\prime)\biggr)\, ,
\label{zbb_mix}
\end{equation}
where $x=m_t^2/M_W^2$, $x^\prime=M_T^2/M_W^2$ and the SM contribution has been subtracted.
In the limit $x,x^\prime>>1$,
\begin{eqnarray}
f_1(x,x^\prime)&=&
x^\prime -x+3\log\biggl({x^\prime\over x}\biggr)
\nonumber \\
f_2(x,x^\prime)&=&
-x-x^\prime +{2 x x^\prime\over x^\prime-x}
\log\biggl({x^\prime \over x}\biggr)
\, .
\label{f1def}
\end{eqnarray}

\section{Numerical Restrictions on VLQs}
\label{globsec}
The properties of VLQ models are restricted by 
$Zb {\overline  b}$, oblique
parameter, and Higgs coupling measurements.  In this section, we perform global
fits to  $Zb {\overline b}$ and oblique
parameter data and demonstrate that Higgs coupling measurements are not competitive
with the  limits from the electroweak parameters.

The  experimental 
constraints from the right-handed $Zb{\overline{b}}$ coupling are
considerably weaker than  those from the left-handed coupling,
so we consider only right-handed couplings, $\delta X_{bb}^R$, that arise at tree level. On the other hand, if
the left-handed coupling is zero at tree level, %when appropriate 
we include the loop corrections 
from $t - T$ mixing 
using  the $1-$loop results of 
Refs.~\cite{Burgess:1993vc,Anastasiou:2009rv} reported in Eq.~\ref{zbb_mix}. 
In the VLQ models where $\delta X_{bb}^R=0$ at tree level, ($T$ and 
$B$ singlet, $(XT)$ doublet, $(TBY)$ triplet), we use the 
$3-$parameter fit to $\Delta S$, $\Delta T$ and $\delta X_{bb}^L$ from
Ref.~\cite{deBlas:2016ojx}\footnote{$\delta X_{bb}^L= 2\delta g_L^b$ 
in the notation of Ref.~\cite{ deBlas:2016ojx}.}.
In addition, $\delta X_{bb}^R\sim 0$ at tree level also  in 
the $(XTB)$ triplet model in the limit $m_b << M_B$~(Eq.~\ref{angrels}), 
and it vanishes in the $(TB)$ doublet model 
when we fix $s_R^b=0$.
In all these cases we use the $3-$parameter fit,
\begin{eqnarray} 
\Delta S&=& 0.10\pm 0.09
\nonumber \\
\Delta T &=& 0.12\pm 0.07\nonumber \\
\delta X_{bb}^L&=& -0.0002\pm 0.0012
\label{3param}
\end{eqnarray}
with the correlation matrix,
\begin{equation}
\rho=\left(
\begin{array}{ccc}
1.0 & 0.85 &0.07 \\
0.85 & 1.0& 0.13\\
0.07 & 0.13 & 1.0
\end{array}
\right)\, .
\label{3cor}
\end{equation}

In the $(BY)$ model we have non-zero
values for 
$\Delta S$, $\Delta T$ and  $\delta X_{bb}^R$.
In the small bottom-mass limit 
$\delta X_{bb}^L$ will be extremely suppressed 
and one can neglect it. Indeed, for $m_b \to 0$
the left-handed $Zb\bar{b}$ coupling is zero at tree level. 
The one-loop contributions vanish as well, since
all the electroweak couplings of the bottom quark are 
proportional to $s_L^b$, which in this limit goes to
zero~(Eq.~\ref{angrels}). 
The $3-$parameter fit
we use in this case is, 
\begin{eqnarray}
\Delta S&=& 0.08\pm 0.09\nonumber \\
\Delta T &=& 0.10\pm 0.07\nonumber \\
\delta X_{bb}^R&=&0.008\pm 0.006
\label{3param_by}
\end{eqnarray}
with the correlation matrix,
\begin{equation}
\rho=\left(
\begin{array}{ccc}
1.0 & 0.86 & -0.19   \\
0.86 & 1.0&  -0.21 \\
-0.19 & -0.21  & 1.0
\end{array}
\label{3cor_by}
\right)\, .
\end{equation}

When both $\delta X_{bb}^L$ and $\delta X_{bb}^R$ are non-zero, we use the $4-$parameter fit of 
Ref.~\cite{deBlas:2016ojx} to
 $\Delta S$, $\Delta T$, $\delta X_{bb}^R$ and $\delta X_{bb}^L$.   For a massless $b$ quark,
 this case only occurs in the $(TB)$ doublet model, where $\delta  X_{bb}^L$ arises at one loop.
 The $4-$parameter fit is, 
\begin{eqnarray}
\Delta S&=& 0.04\pm 0.09\nonumber \\
\Delta T &=& 0.08 \pm 0.07\nonumber \\
 \delta X_{bb}^L&=& 0.006 \pm 0.002\nonumber \\
\delta X_{bb}^R&=&0.034\pm 0.016
\label{4param}
\end{eqnarray}
with the correlation matrix,
\begin{equation}
\rho=\left(
\begin{array}{cccc}
1.0 & 0.86 & -0.24 &-0.29   \\
0.86 & 1.0& -0.15 & -0.22 \\
-0.24 & -0.15 & 1.0 & 0.91 \\
-0.29 & -0.22 & 0.92 & 1.0
\end{array}
\label{4cor}
\right)\, .
\end{equation}
We perform a $\chi^2$ fit,
\begin{equation}
\Delta \chi^2=\Sigma_{ij}(O_i-O_i^{fit})(\sigma^2)^{-1}_{ij}(O_j-O_j^{fit})\, ,
\end{equation}
where $O_i$ are the measured observables ($\Delta S$, $\Delta T$,
$\delta X_{bb}^L$, $\delta X_{bb}^R$),
$O_i^{fit}$ are their predicted values in the different VLQ models, 
and $\sigma^2_{ij}=\sigma_i\rho_{ij}\sigma_j$, where
$\sigma_i$ are the uncertainties  in Eqs.~\ref{3param},~\ref{3param_by}
and~\ref{4param}. The correlation matrices are given in 
Eqs.~\ref{3cor},~\ref{3cor_by} and \ref{4cor}.  
In each model, we scan over the parameters to obtain the  $95\%$
confidence level limits. 
All the models but the $(TB)$ doublet have two independent degrees of
freedom (see Eq.~\ref{pardef}). Also in the $(TB)$ model we will analyse two specific scenarios,
one with $s_R^b$ fixed, and one with $M_T$ fixed. Therefore, in all
cases the number of degrees of freedom is two, and we
require $\Delta \chi^2<5.99$.

\begin{figure}[tb]
\centering
\includegraphics[width=0.45\textwidth,clip]{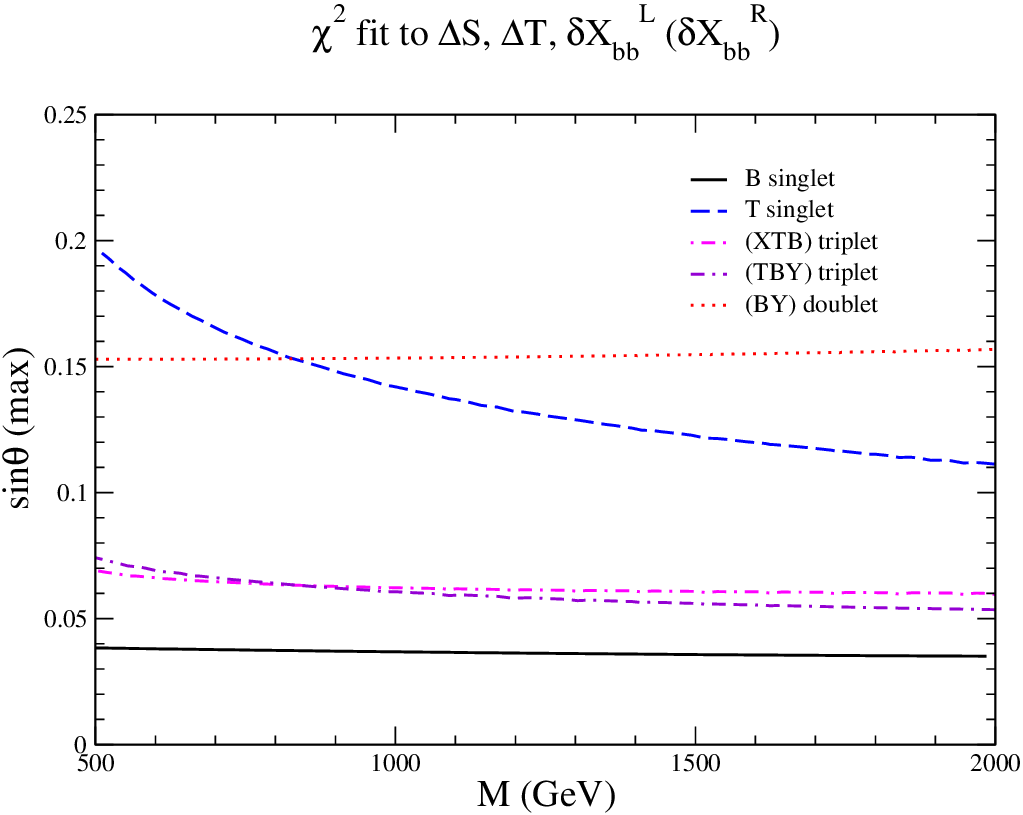}
\caption{$95\%$ confidence level allowed regions in the various VLQ models models, where
the meanings of $\sin\theta$ and $M$ are defined in Eq.~\ref{pardef}. The regions below the curves
are allowed.}
\label{fig:fit2}
\end{figure}

\begin{figure}[tb]
\centering
\includegraphics[width=0.45\textwidth,clip]{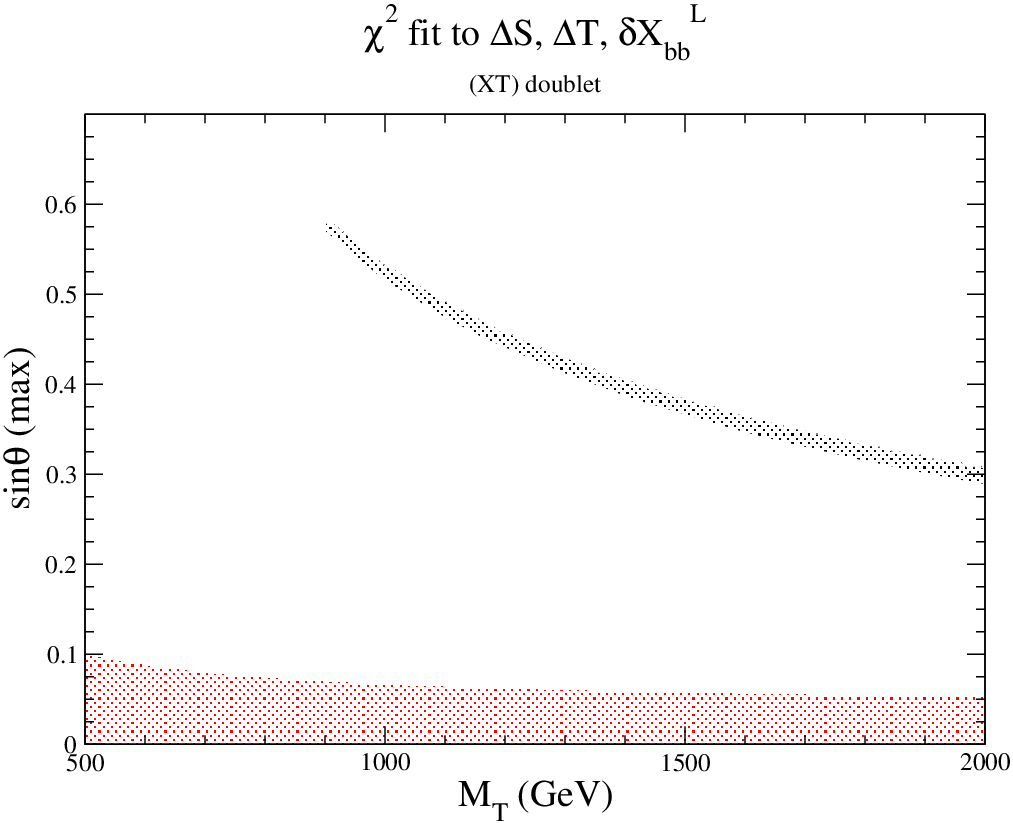}
\caption{$95\%$  confidence level allowed regions in the $(XT)$  doublet model, where
the dotted regions
are allowed.}
\label{fig:fitxt}
\end{figure}

\begin{figure}[tb]
\centering
\includegraphics[width=0.45\textwidth,clip]{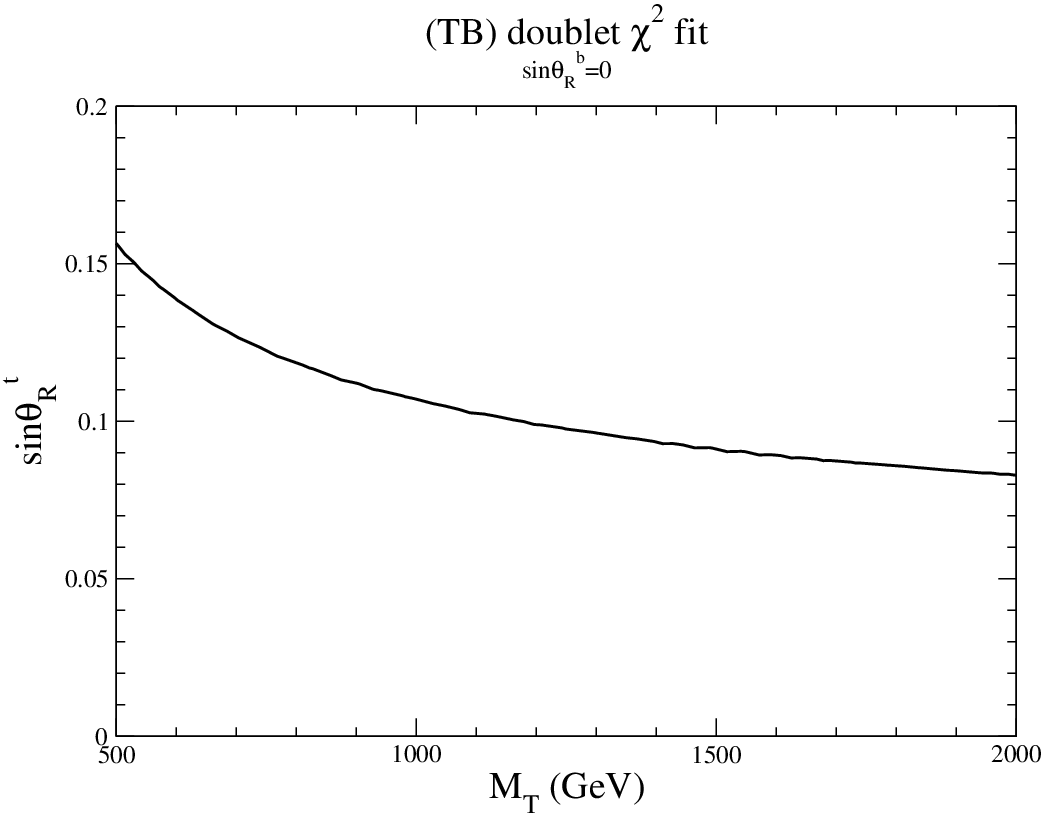}
\includegraphics[width=0.45\textwidth,clip]{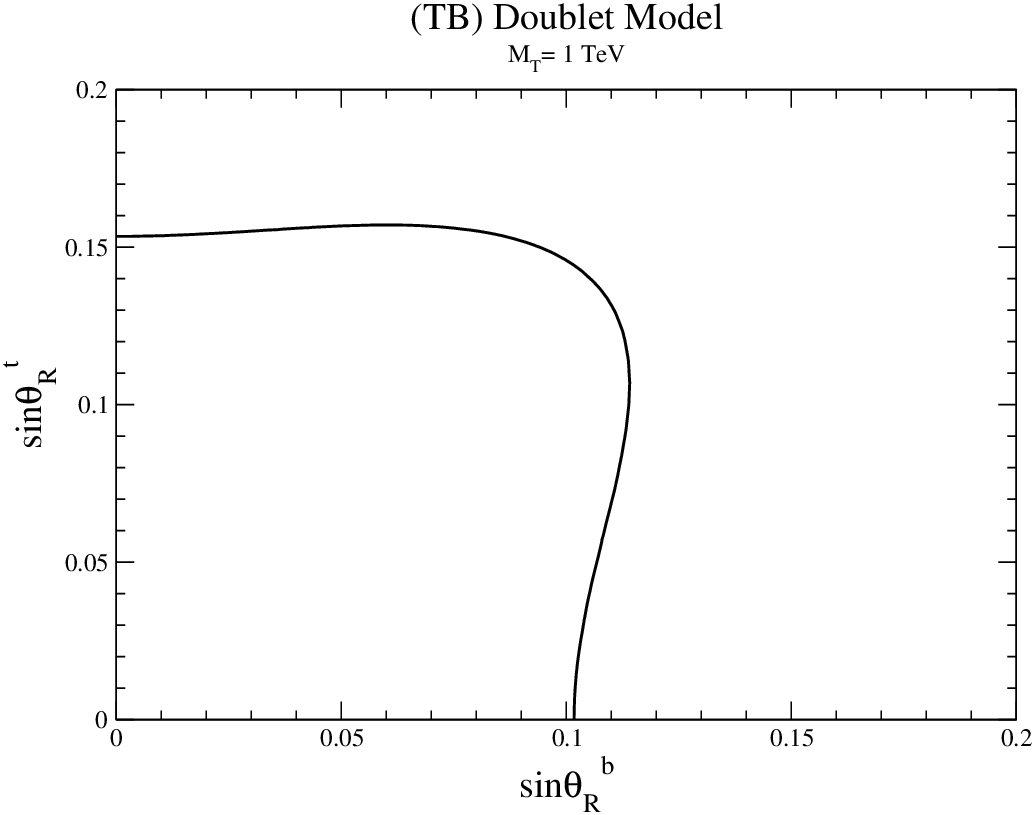}
\caption{$95\%$ confidence level allowed regions in the $(TB)$ doublet model
for  $s_R^b=0$ (LHS) and $M_T=1~$TeV (RHS).
The regions below the curves
are allowed.}
\label{fig:fittb}
\end{figure}

Our results for the regions of parameter space allowed by the electroweak precision
observables are shown in Figs.~\ref{fig:fit2},~\ref{fig:fitxt} and \ref{fig:fittb},  where we use the exact results 
for the oblique parameters\footnote{The exact result for $\Delta S$ 
in the $(XT)$ doublet model shows numerical  instabilities in the small angle region.
Hence, we have used an expansion up to ${\cal O}\left((s_R^t)^{16}\right)$ 
for $s_R^t<$  0.2. At the matching point, the exact result is stable and the
difference with the expanded one is below the percent level.}. 

Ref.~\cite{Aguilar-Saavedra:2013qpa} showed limits from the oblique parameters and the 
$Zb{{\overline b}}$ couplings separately, and our fit
results are roughly consistent with theirs, although the experimental constraints have 
tightened somewhat.  For the $T$, $B$  singlet models, the $(BY)$ doublet model and the triplet
models, the limits on the mixing angles are quite stringent and for large VLQ masses
relatively independent of the VLQ mass itself (Fig.~\ref{fig:fit2}).

The $(XT)$ doublet model has an interesting region  seen in Fig.~\ref{fig:fitxt} (black dotted
area), where the contribution
to $\Delta T$ vanishes, allowing  relatively large values of the
mixing angle.  This region is consistent with the $\Delta T \sim 0$ 
region of Fig.~\ref{fig:dt} for $M_T=1~$TeV.
\begin{figure}[tb]
\centering
\includegraphics[width=0.6\textwidth,clip]{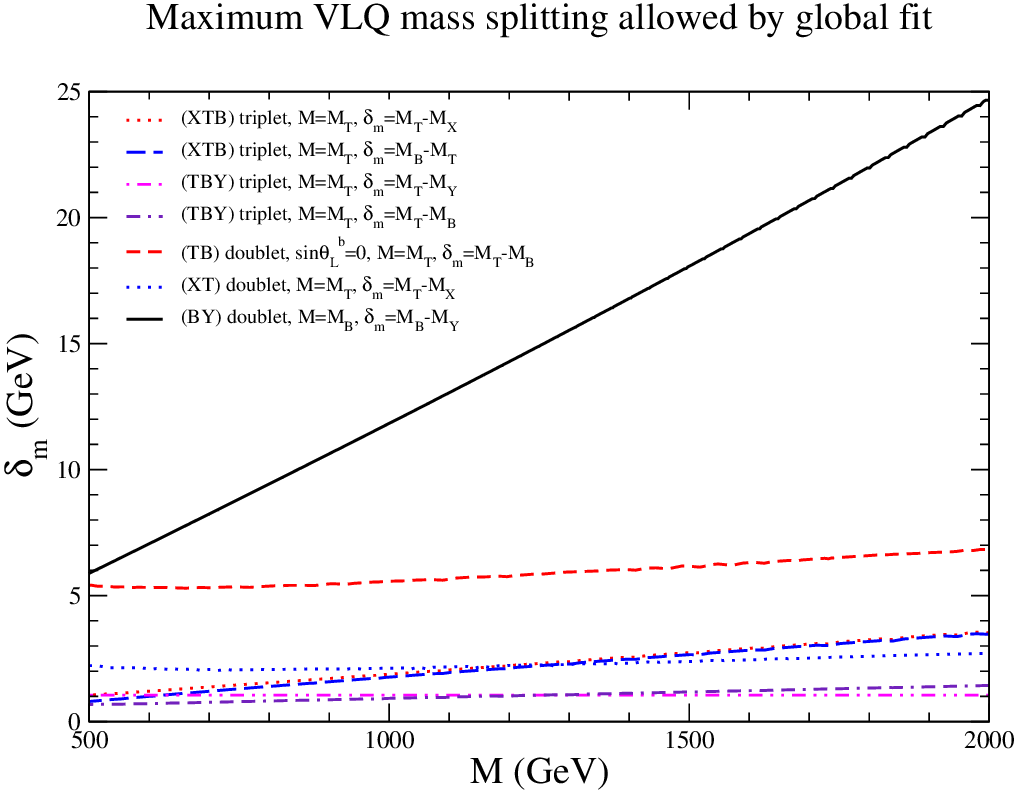}
\caption{Maximum allowed mass splitting between the members of a VLQ
multiplet using the results of Fig.~\ref{fig:fit2}.}
\label{fig:fitmass}
\end{figure}
\\
Also in the $(TB)$ doublet model (Fig.~\ref{fig:fittb}) we find an interesting 
region with relative large mixings both in the top and bottom sectors 
allowed by the fit.  
Doing a global fit strengthens the bounds in the $(TB)$  and $(XTB)$ models relative to those of
Ref.~\cite{Aguilar-Saavedra:2013qpa}. 
Models with $B$ VLQs~\cite{Gillioz:2013pba} are allowed by the fits,  with a  relatively
large mixing angle permitted in the $(BY)$ doublet model.  
The strongest limit on models with $B$ VLQs  occurs in the $B$ singlet case,
where for all $M_B$,  the global fit requires $s_L^b<0 .04$ due to the
strong dependence of $\delta X^L_{bb}$ on the mixing angle (Eq.~{\ref{xbbdef}).
 We note that for large VLQ masses, the fits asymptote to an approximately  constant
mixing angle in each case.  This suggests that the value of the VLQ mass is not
critical and that an effective field theory (EFT) approach is warranted.  We discuss the EFT approach for heavy
VLQs in Appendix B.   

We have presented our results in terms of the masses and mixing angles given in Eq.~\ref{pardef}.  Using
Eq.~\ref{sigrels}, we redisplay our fit results in terms of the allowed mass differences between 
members of the VLQ multiplets.  In Fig.~\ref{fig:fitmass}, 
we demonstrate that the maximum allowed mass differences
are of ${\cal{O}}(1-3~{\rm GeV})$, except for the $(BY)$   and $(TB)$ doublet models,
where mass differences of  ${\cal{O}}(5-10~{\rm GeV})$ are consistent with precision electroweak data.

The mixing of SM and vector-like quarks also changes the $tbW$ couplings, $A_{tb}^{L,R}$.
The  limits from 7~and~8~TeV data from $t-$channel single top
production~\cite{Khachatryan:2014iya,Aguilar-Saavedra:2013qpa} are
however not yet competitive with the precision electroweak limits.

Finally, in the VLQ models Higgs production and decay rates are
modified. The Higgs signal strengths for the gluon fusion 
production channel using the  $95\%$ confidence
level results at 8~TeV are~\cite{ATLAS-CONF-2015-044},
\begin{eqnarray}
\mu_F^{\gamma\gamma}&=&1.13^{+0.24}_{-0.21}\nonumber \\
\mu_F^{WW}&=&1.08^{+0.22}_{-0.19} \nonumber \\
\mu_F^{ZZ}&=&1.29^{+0.29}_{-0.25} \nonumber \\
\mu_F^{bb}&=&0 .65 ^{+0.37}_{-0.28}\nonumber \\
\mu_F^{\tau\tau} &=& 1.07^{+0.35}_{-0.28}\, .
\label{eq:mulims}
\end{eqnarray}
The production rate $gg\rightarrow h$ and the decays $h\rightarrow gg$ and
$h\rightarrow \gamma\gamma$ are affected by the VLQ contributions 
through loops of heavy quarks and changes in the SM quarks Yukawa 
couplings, while the $h\rightarrow b\bar{b}$ decay is modified at tree
level. 

The contribution to the Higgs signal strength from colored fermions is
well known. At leading order~\cite{Gunion:1989we},
\begin{equation}
\mu_{ggF}\equiv {\sigma(gg\rightarrow h)\over \sigma(gg\rightarrow h)\mid_{SM}}
={\mid \sum_{f=t,b, T,B} {c_{ff} \over m_f} F_F(\tau_f)\mid^2\over
\mid \sum_{f=t,b}  {c_{ff}^{SM} \over m_f} F_F (\tau_f)\mid^2}\, ,
\end{equation}
where $c_{ff}$ are the Higgs-fermion couplings defined in
Eq.~\ref{higgscoup}, $m_f$ is the mass of the corresponding quark, 
$c_{ff}^{SM}=m_f$, $\tau_f={m_h^2\over 4M_f^2}$,
and
\begin{eqnarray}
F_F(x)&=& {2\over x^2}\biggl[x+(x-1)f(x)\biggr]\nonumber \\
f(x)&=&
\left\{
\begin{matrix}
 \biggl[ \sin^{-1}(\sqrt{ x})\biggr]^2\qquad \qquad\quad\;\; x< 1\nonumber \\
-{1\over 4} \biggl[\ln(x_+/x_-)-i \pi\biggr]^2\qquad x>1
\end{matrix} \right. \nonumber \\
x_\pm&=& 1\pm \sqrt{1-{1\over x}}\, . 
\label{eq:FFxdef}
\end{eqnarray}
In the heavy fermion mass limit $F_F(x)\rightarrow {4\over 3}$, while 
for light quarks $F_F(\tau_b)\rightarrow 0$. 
Therefore in the  limit of massless $b$ quark and infinitely heavy
$(t,T,B)$ quarks, 
the leading order Higgs production rate is independent of the
fermion masses,
\begin{equation}
\mu_{ggF}= {\sigma(gg\rightarrow h)\over \sigma(gg\rightarrow h)\mid_{SM}}
\rightarrow \,\mid \!\Sigma_{f=(t,T,B)} {c_{ff} \over m_f}\mid^2 \, .
\end{equation}

The deviations of the
gluon fusion  production rate,
$\mu_{ggF}$,
are directly related to deviations in the $b$ couplings,
\begin{eqnarray}
T, (XT):& \mu_{ggF}\rightarrow & 1\nonumber \\
B, (TBY):& \mu_{ggF}\rightarrow & 1+2 (s_L^b)^2 = 1+\delta X_{bb}^L\nonumber \\
(XTB):& \mu_{ggF}\rightarrow & 1+2 (s_L^b)^2 = 1-\delta X_{bb}^L\nonumber \\
(TB): & \mu_{ggF}\rightarrow & 1+2 (s_R^b)^2 = 1-\delta X_{bb}^R\nonumber \\
(BY): & \mu_{ggF}\rightarrow & 1+2 (s_R^b)^2 = 1+\delta X_{bb}^R \;.
\end{eqnarray}
% Comparing with the definitions of $\delta X_{bb}$ in
% Eq.~\ref{xbbdef}, 
We observe 
that in all cases the presence of heavy $B$ VLQs ${\it{increases}}$ the Higgs signal strength.

For the decay width to photons  we have,
\begin{eqnarray}
\mu^{\gamma\gamma}&\equiv &{\Gamma(h\rightarrow \gamma\gamma)
\over \Gamma(h\rightarrow \gamma \gamma)\mid_{SM}}\nonumber \\
&=&
{\mid \Sigma_{f=f_{SM}, T,B}N_{c}Q_f^2 (c_{ff}/m_f) F_F(\tau_f)+F_W(\tau_W)\mid^2\over
\phantom{aA}\mid \Sigma_{f=f_{SM}} N_{c}Q_f^2 F_F(\tau_f)+F_W(\tau_W) \mid^2}\, ,
\end{eqnarray}
where $f_{SM}$ includes all SM fermions, $Q_f$ and $N_{c}$ are charge
and color of the fermion, $F_F(x)$ is defined in Eq.~\ref{eq:FFxdef} and
\begin{eqnarray}
F_W(x)& =&-{1\over x^2}\biggl[ 2x^2+3x+3(2x-1)f(x)\biggr] \qquad{\rm ,
           \quad with}\nonumber \\
F_W(0)&\rightarrow& -7 \;.
\end{eqnarray}
Modifications of Higgs signal strength for the various VLQs are shown in
Figs.~\ref{fig:hfit1},~\ref{fig:hfit_bsing},~\ref{fig:hfit_ttb} and~\ref{fig:hfit_trip}, where we define,
\begin{equation}
\mu_{ggF}^{XX}\equiv {\sigma(gg\rightarrow h)BR(h\rightarrow XX)\over
[\sigma(gg\rightarrow h)BR(h\rightarrow XX)]_{SM}} \;.
\end{equation}

The $B$ and $T$ singlet and $(BY)$ $XT$ doublet models are so highly constrained by the electroweak fits,
that the deviations in Higgs production are too small to be observed. 
The $(TB)$ doublet model can have modest increases in Higgs signal 
strengths when the mixing in the right-handed $b$ sector is allowed to be significant
(RHS of Fig.~\ref{fig:hfit_ttb}). 
 In the $(XTB)$ triplet triplet model, an increase of about $10\%$ in the $h\rightarrow \gamma
\gamma$ signal strength is consistent with the results from the electroweak fits, while
the other Higgs decay channels are constrained to be within about $4\% $  of the SM
predictions. 

\begin{figure}[tb]
\centering
\includegraphics[width=0.45\textwidth,clip]{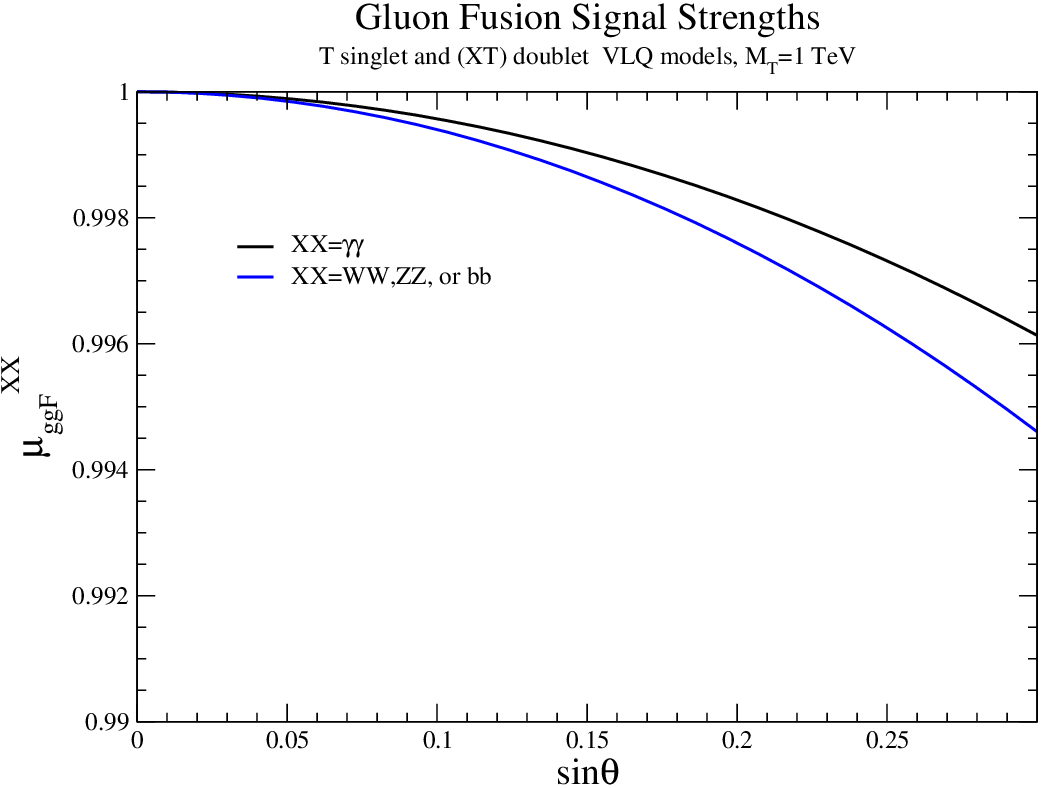}
\caption{Higgs branching ratios in VLQ models with a $(T)$ singlet
or an $(XT)$ doublet normalized to the Standard Model predictions. }
\label{fig:hfit1}
\end{figure}

\begin{figure}[tb]
\centering
\includegraphics[width=0.45\textwidth,clip]{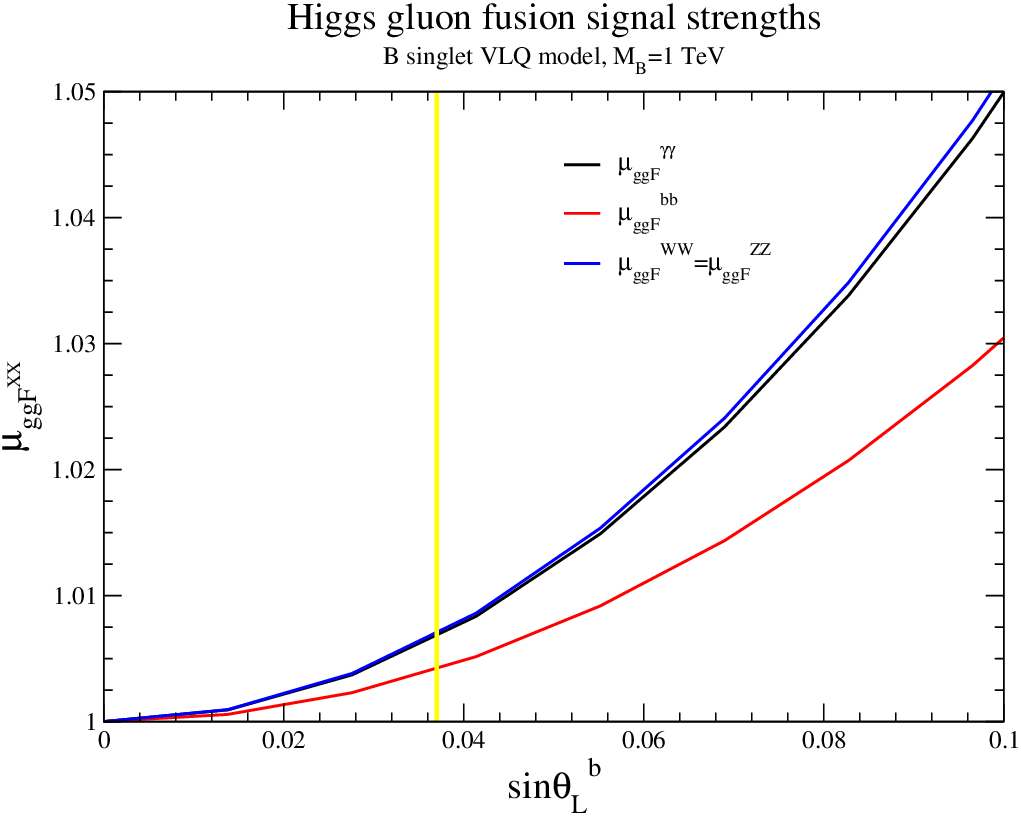}
\includegraphics[width=0.45\textwidth,clip]{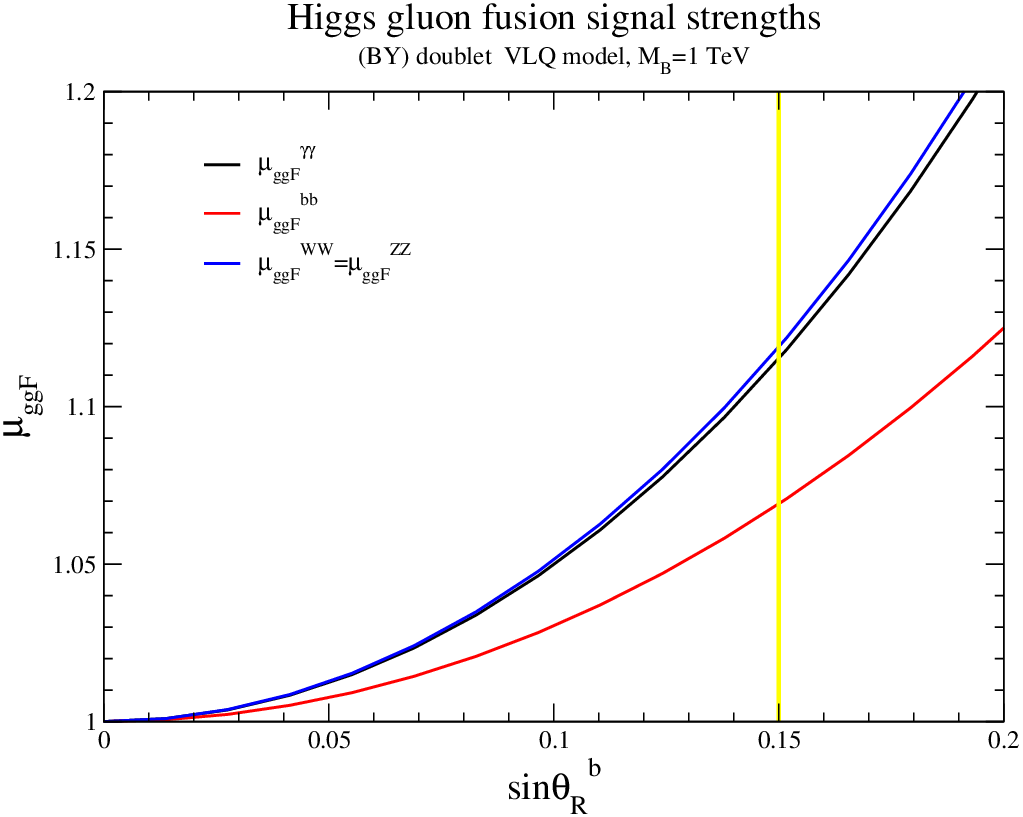}
\caption{Higgs branching ratios in VLQ models normalized to the SM predictions. The
vertical yellow lines are the maximum mixing allowed by the electroweak
fits for the $B$ singlet (LHS) and $(BY)$ doublet (RHS)  models shown in Fig.~\ref{fig:fit2}. }\label{fig:hfit_bsing}
\end{figure}

\begin{figure}[tb]
\centering
\includegraphics[width=0.45\textwidth,clip]{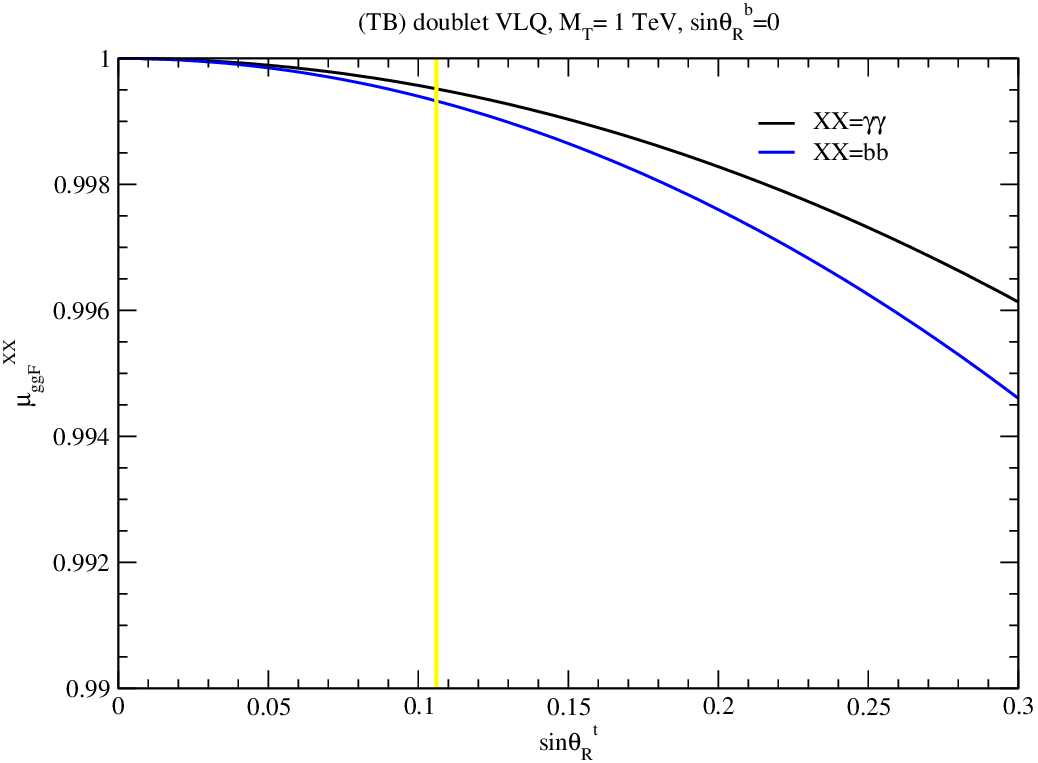}
\includegraphics[width=0.45\textwidth,clip]{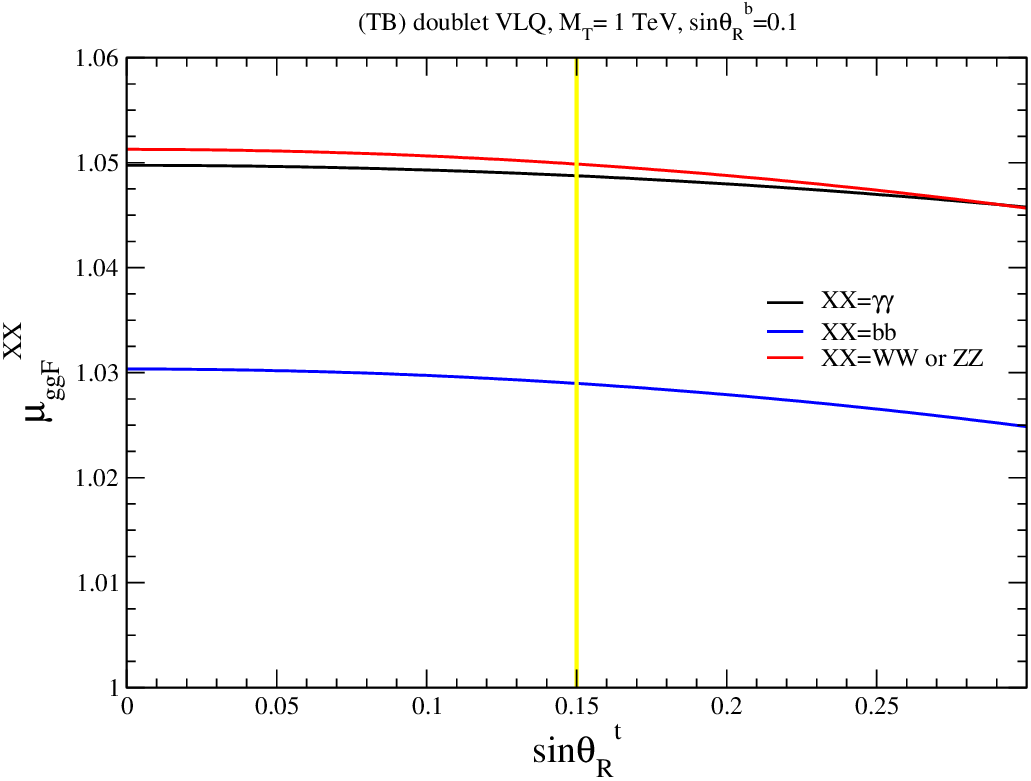}
\caption{Higgs gluon fusion signal strengths in $(TB)$ doublet VLQ model, normalized to the SM
predictions.  RHS has $s_R^b=0.1$ and LHS has $s_R^b=0$. The vertical yellow lines are the maximum mixing allowed by the electroweak fits
of Fig.~\ref{fig:fittb}.}
\label{fig:hfit_ttb}
\end{figure}

\begin{figure}[tb]
\centering
\includegraphics[width=0.45\textwidth,clip]{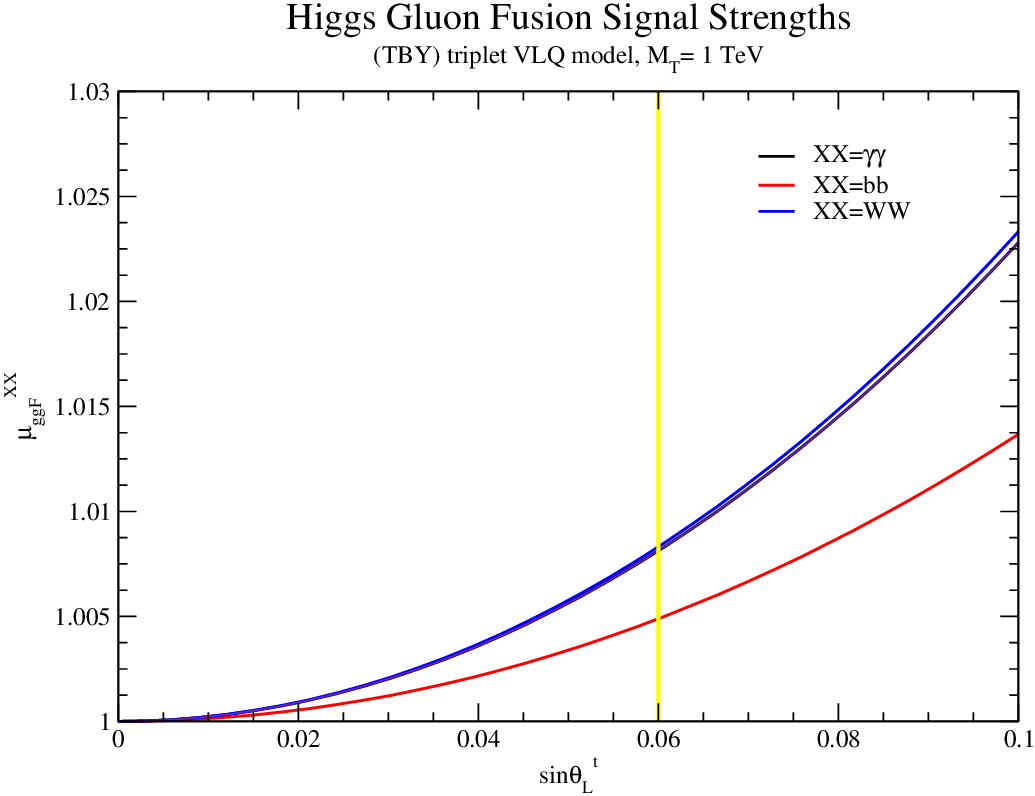}
\includegraphics[width=0.45\textwidth,clip]{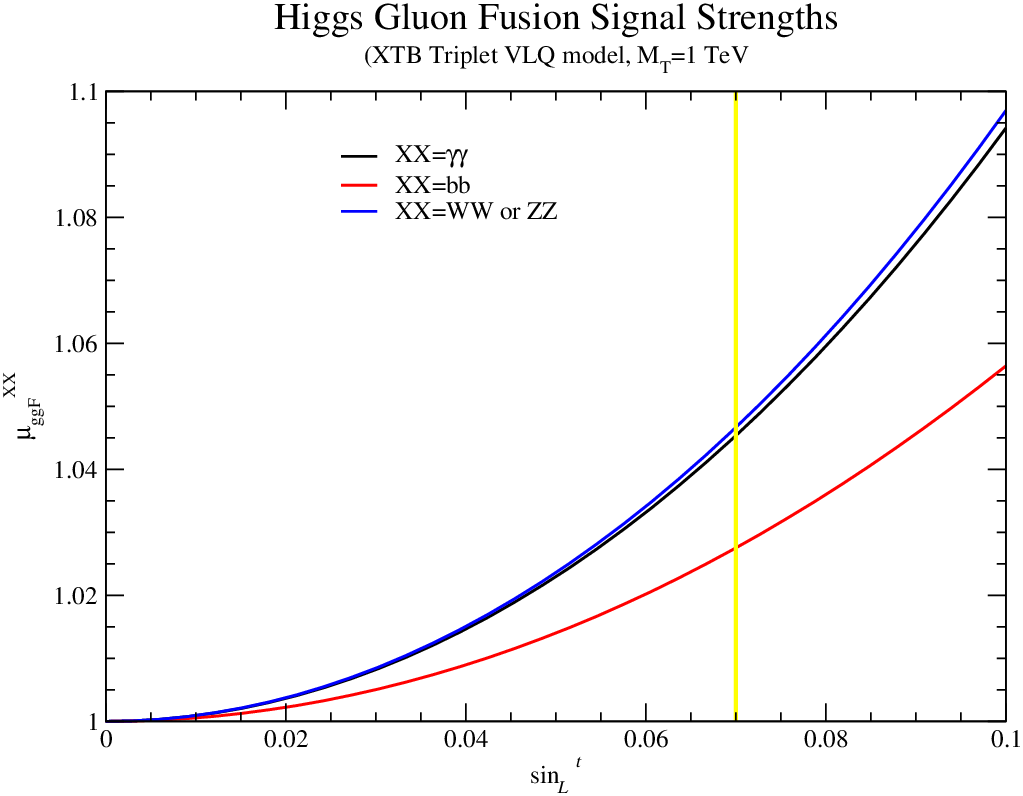}
\caption{Higgs branching ratios in the VLQ models normalized to the Standard Model predictions. The
vertical yellow lines are the maximum mixing allowed by the electroweak
fits for the $TBY$ triplet (LHS) and $(XTB)$ triplet (RHS)  models shown in Fig.~\ref{fig:fit2}.}
\label{fig:hfit_trip}
\end{figure}

\section{conclusions}
\label{consec}
We have considered restrictions on the parameters of models with vector-like quarks and 
updated electroweak fits to the parameters of these models.  
The constraints on VLQ masses and mixings are strengthened from previous
fits.  Mixing in the $B$ VLQ sector is highly constrained
due to the tree-level effect on the $Zb\bar{b}$ coupling, 
while mixings up to $s_R^t \sim 0.2$ are allowed in the $T$ singlet case.  
In the doublet models mixings up to $s_R \sim 0.1 - 0.15$ are allowed, 
with an interesting region of $0.3 \lesssim s_R^t \lesssim 0.6$ in the 
$(XT)$ doublet scenario and 
non-zero mixing allowed in both the top and bottom sectors
in the $(TB)$ doublet model. 
In the triplet models the mixing is somewhat more constrained, $s_L^t \lesssim 0.7$. 
Finally, we show that in order for Higgs coupling measurements to probe regions beyond those excluded
by precision fits, measurements of a few $\%$ will be required.

\section*{Acknowledgements}
We would like to thank G. Panico for useful discussion. 
SD thanks B. Jaeger and the University of Tuebingen where this work was begun
and Fermilab where it was finally completed for their hospitality. 
 This work is supported by the U.S. Department of Energy under grant
 DE-SC0012704, by the Swiss National Science Foundation (SNF) under contracts 
200021-165772 and 200021-160814, and by the Advanced ERC Grant Pert QCD.
The work of C.-Y.C is supported by NSERC, Canada.
Research at the Perimeter Institute is supported in part by the Government of Canada through NSERC and by the Province of Ontario through MEDT.
Digital data related to our results can be found at~
\href{$quark.phy.bnl.gov\backslash Digital\_Data\_Archive\backslash dawson\backslash vlq\_17$}{https://quark.phy.bnl.gov/Digital_Data_Archive/dawson/vlq_17/}

\renewcommand{\theequation}{\thesection.\arabic{equation}} 

\begin{appendix}
\section{VLQ Triplet Lagrangian}
\label{VLTapp}
In the following, the fields are the current eigenstates, but for simplicity
of notation we shall omit the superscript ``0".\\
To establish our normalization convention, we shall use
\beq
	\lag^{(SM)} 
	=
	\overline{\psi}_L \gamma^\mu i \left(
		\partial_\mu	 - i g W_\mu(x)-i g' 1\!\! 1_{(2)} B_\mu(x)\right)
		\psi_L 
		+
		\overline{q}_R \gamma^\mu i \left(\partial_\mu	-i g' 1\!\! 1_{(1)} B_\mu(x)\right)
		q_R \;,
\eeq
where $\psi_L$ is the top-bottom left-handed doublet, 
$q_R = \{t_R, b_R\}$ are the right-handed singlets,
$W_\mu = W_\mu^a \frac{\sigma^a}{2}$ ($a=\{1,2,3\}$) and $\sigma^a$  are the Pauli matrices. 
They satisfy 
\beq
	\left[ \sigma^a , \sigma^b\right] = 2 i \epsilon^{abc}\sigma^c \qquad,\quad
	{\rm Tr}[\sigma^a \sigma^b] = 2 \delta^{ab} \;.
\eeq
The gauge boson interactions therefore read
\bea
\label{eq:couplingsSM}
	\lag^{(SM)}_{g.b.} &=&	
			e A_\mu 
			\left[
			 \bar{t}_{L,R}  \gamma^\mu  \left(
			 	\frac{I^{3(SM)}_{t_{L,R}}}{2}+Y_{t_{L,R}}^{(SM)}\right) t_{L,R} 
			 + \bar{b}_{L,R}  \gamma^\mu  \left(
			 	\frac{I^{3(SM)}_{b_{L,R}}}{2}+Y_{b_{L,R}}^{(SM)}\right) b_{L,R}
			\right] + \nn
			&& \frac{g}{2 c_W} Z_\mu
			\left[
				\bar{t}_{L,R} \gamma^\mu \left(I^{3(SM)}_{t_{L,R}} - 2 Q^{(SM)}_t s_W^2 \right) t_{L,R}
				+ \bar{b}_{L,R}\gamma^\mu  \left(
					I^{3(SM)}_{b_{L,R}} -2 Q^{(SM)}_b s_W^2\right) b_{L,R}
			\right]	+ \nn
			&& \frac{g}{\sqrt{2}}  W_\mu^+ \bar{t}_L \gamma^\mu b_L 
			+\frac{g}{\sqrt{2}} W_\mu^- \bar{b}_L \gamma^\mu t_L \;,
\eea
%%where we used %, for a quark doublet $\psi$, 
%%\beq
%%	\sigma^3 \psi_L = I^3_q \psi_L \qquad,\quad 1\!\!1_{(2)} \psi = Y_q^{(SM)} \psi \;,
%%\eeq
%%with\footnote{We denote $SU(2)$ singlets and doublets by the superscripts
%%$s$ and $d$ respectively.}
%%and the charges are
%
where $I^{3(SM)}_q, Y^{(SM)}_q$ are the eigenvalues of the quark $q=\{t_{L,R},b_{L,R}\}$ under
$\sigma_3$ and the $U(1)_Y$ generator respectively,
\bea
	I^{3(SM)}_{t_R} = I^{3(SM)}_{b_R} = 0
	\phantom{11111222} &\qquad,\quad&
	I^{3(SM)}_{t_L} =- I^{3(SM)}_{b_R} = 1 \nonumber\\
	Y^{(SM)}_{t_R} = \frac{2}{3} \quad,\quad
	Y^{(SM)}_{b_R} = -\frac{1}{3} &\qquad,\quad&
	Y^{(SM)}_{\psi_L} = \frac{1}{6}   \;.
\eea
This yields the correct charges for the top and bottom quarks (first line of eq.~(\ref{eq:couplingsSM})). 
In the second line we replaced the hypercharge quantum
number with the corresponding charge of the quark as derived from
the first line. 

As a reminder, the physical gauge bosons are defined by 
\beq
		W^{\pm} = \frac{W^1 \mp i W^2}{ \sqrt{2}} \qquad,\quad
		W^3_\mu = c_W Z_\mu + s_W A_\mu  \qquad,\quad 
		B_\mu = c_W A_\mu - s_W Z_\mu \;,
\eeq
and $ e = g s_W = g' c_W$.

\subsection{Vector Triplets}

\section*{Introducing the fields}
Let $\rho^0$ be a fermionic field that transforms as a triplet under
$SU(2)_L$, i.e.\footnote{
	Recall that the triplet vector field $W_\mu$ transforms as 
	$$
		W_\mu \rightarrow U \left( W_\mu +\frac{i}{g} \partial_\mu\right) U^{-1} \;.
	$$
	A triplet fermionic field has a similar transformation law, up to the 
	shift term.}
\beq
	\rho^0 \rightarrow U(x) \rho^0 U^{-1}(x) \qquad \quad , \qquad
	U(x) = e^{i \alpha^a(x) \tau^a} \;.
\eeq
Here $\alpha^a(x)$ are the gauge transformation parameters and 
$\tau^a = \frac{\sigma^a}{2}$.
%, and the commutator of the Pauli matrices $\sigma^i$ is
%\beq
%	[\sigma^i,\sigma^j] = i \varepsilon_{ijk} \sigma^k 
%	\rightarrow [\tau^i,\tau^j] = \frac{i}{2} \varepsilon_{ijk} \tau^k 
%	\nonumber \;.
%\eeq
For simplicity of notation, from now on we will omit the subscript``0" in 
the fermionic fields and use $U \equiv U(x)$. \\
Note that we can decompose fermions in a $SU(2)_L$ triplet 
on the basis of Pauli matrices as 
\beq
	\rho = \rho^i \tau^i \qquad (W_\mu =  W_\mu^i \tau^i) \;.
\eeq
As for the charged gauge bosons, we introduce 
\beq
	\rho^{\pm} = \frac{\rho^1 \mp i \rho^2}{ \sqrt{2}} \;,
\eeq
and we can use that 
\beq
	\rho^1 \tau^1 + \rho^2 \tau^2 
	= 
	\rho^+ \frac{\tau^1 + i \tau^2}{\sqrt{2}} + \rho^- \frac{\tau^1 - i \tau^2}{\sqrt{2}} \;.
\eeq
Let us remind ourselves that the charges of $ \tau^1 \pm i \tau^2$ (i.e. 
$\rho^\pm$ and $W^\pm$) are $\pm 1$ respectively,
\beq
	[\tau^3, \tau^1 \pm i \tau^2] = \pm (\tau^1 \pm i \tau^2) \;.
\eeq

\section*{Guage invariance and the Lagrangian}
From the transformation laws of the gauge and fermion fields, 
the  gauge-invariant covariant derivative must be defined 
as\footnote{We want $D_\mu \rho \rightarrow U \left( D_\mu \rho \right) U^{-1}$, 
and we need to pick the same normalization conventions as those for the Standard Model.}
\beq
	D_\mu \rho = \partial_\mu \rho - i  \, [W_\mu,\rho] \;.
\eeq
%The proof goes along the same lines as in 
%Eqs.~(\ref{eq:def_cov_deriv})--(\ref{eq:def_cov_deriv_SM}).
The Lagrangian, imposing also the correct normalization of the 
kinetic term and adding the $U(1)_Y$ part, is then
\bea
   \label{eq:lag_triplet}
	\lag & = & 2 \textrm{Tr} \left\{\bar{\rho} \, i \gamma^\mu D_\mu \, \rho \right\}
				+ 2 \textrm{Tr} \left\{g' Y
					\bar{\rho} \gamma^\mu B_\mu \rho \right\} \\
		& = & 
			\frac{1}{2} \textrm{Tr} \left\{\sigma^a \sigma^b\right\}
				i \bar{\rho}^a \gamma^\mu (\partial_\mu\rho^b)
			+ \frac{1}{2} \textrm{Tr} \left\{ \sigma^a \left[\sigma^b, \sigma^c\right] \right\}
					\left(\frac{g}{2} \bar{\rho}^a \gamma^\mu W_\mu^c \rho^c\right)
			+ \frac{1}{2} \textrm{Tr} \left\{\sigma^a \sigma^b\right\} 
					\left(g' Y \bar{\rho}^a \gamma^\mu B_\mu\rho^b\right)
		\nn
		& = &  \bar{\rho}^a \gamma^\mu (\partial_\mu \rho^a)
			+ ig \bar{\rho}^a \gamma^\mu W_\mu^c \rho^c \epsilon^{abc} 
			+ g' Y \bar{\rho}^a \gamma^\mu B_\mu\rho^a
 \;. \nonumber
\eea
%%%%The rest of the calculation is carried out in the notes and in the Mathematica notebook
%%%%\texttt{triplet MathematicaX.nb}. One just needs to replace
%%%%\beq
%%%%	f^1 = \frac{f^+ + f^-}{\sqrt{2}} \qquad , \quad f^2 = i \frac{f^+ - f^-}{\sqrt{2}} 
%%%%	\qquad, \quad f= \{ \rho, W_\mu \} \;,
%%%%\eeq
%Note that we keep track of the isospin $T^3_q$, which is 
%$\pm 1$ for $\rho^\pm$ and 0 for $\rho^0$, as we proved above by taking the 
%commutator with $\sigma^3$. 

\section*{Couplings of the fermions to the EW gauge bosons}
Let us recall that in the normalization we chose 
the $\{X,T,B\}$ triplet has isospin
\mbox{$I^3_{\rho_{XTB}}=\{2,0,-2\}$}, \mbox{$Y_{\rho_{XTB}}=\frac{2}{3}$}, 
and the $\{T,B, Y\}$ triplet has \mbox{$I^3_{\rho_{TBY}}=\{2,0,-2\}$}, 
\mbox{$Y_{\rho_{TBY}}=-\frac{1}{3}$}.
The couplings to the electroweak gauge bosons are easily derived from
eq.~(\ref{eq:lag_triplet})
via the replacements 
\beq
	f^1 = \frac{f^+ + f^-}{\sqrt{2}} \qquad , \quad f^2 = i \frac{f^+ - f^-}{\sqrt{2}} 
	\qquad, \quad f= \{ \rho, W_\mu \} \;.
\eeq
%
%\subsubsection*{Standard Model couplings}
%I write them here to establish the normalization
%\bea
%	\lag^{(SM)} &=&		 	
%			\frac{1}{2} e \gamma_\mu 
%			\left[
%			 \bar{t}  \gamma^\mu  \left(T^{3(SM)}_q+Y^{(SM)}\right) t 
%			 + \bar{b}  \gamma^\mu  \left(- T^{3(SM)}_q-Y^{(SM)}\right) b
%			\right] + \nn
%			&& \frac{g}{2 c_W} Z_\mu
%			\left[
%				\bar{t} \gamma^\mu \left(T^{3(SM)}_q - Q^{(SM)}_t s_W^2 \right) t
%				+ \bar{b}\gamma^\mu  \left(-T^{3(SM)}_q -Q^{(SM)}_b s_W^2\right) b
%			\right]	+ \nn
%			&& \frac{g}{\sqrt{2}}  W_\mu \bar{t} \gamma^\mu b 
%			+\frac{g}{\sqrt{2}} W_mu \bar{b} \gamma^\mu W_\mu t \;.
%\eea
%Since $T^{3(SM)}_q=\frac{1}{2}, Y^{(SM)}=\frac{1}{6}$, one recovers the charges 
%of the top and the bottom, $\frac{2}{3}$ and $-\frac{1}{3}$ respectively. The couplings
%to the $Z$ boson read
%\beq
%	\lag^{(SM)}_Z =  \frac{g}{2 c_W} Z_\mu
%			\left[
%				\bar{t} \gamma^\mu \left(\frac{1}{2} - \frac{2}{3} s_W^2 \right) t
%				+ \bar{b} \gamma^\mu \left(-\frac{1}{2}+ \frac{1}{3} s_W^2\right) b
%			\right]	\;.
%\eeq
%%
\subsubsection*{Neutral couplings: the photon}
The photon couplings allow us to determine the electric charge of the three quarks.
From the Lagrangian~(\ref{eq:lag_triplet}), with the definition of the gauge 
boson fields of the Standard Model, one gets
%%\footnote{In my normalization for the Standard Model, there is a factor of 1/2
%%	in front of the charge $Q_q = (T^3_q+Y)$ of a quark $q$, and its hypercharge is
%%	$\frac{g}{2 c_W} (T^3_q - s_W^2 Q_q)$.}
%
\beq
	\lag_\gamma  =  e A_\mu \left[
		\bar{\rho}^+ \gamma^\mu \rho^+ \left(Y+\frac{I^3_{\rho^+}}{2}\right) + 
		\bar{\rho}^3 \gamma^\mu \rho^3 \left(Y\right) + 
		\bar{\rho}^- \gamma^\mu \rho^- \left(Y+\frac{I^3_{\rho^-}}{2}\right) \right] \;.
\eeq
Hence
\begin{itemize}
\item  for the triplet of hypercharge $Y=\frac{2}{3}$, $\rho^+ \equiv X$ has charge 5/3, $\rho^3 \equiv T$ has charge 2/3, 
and $\rho^- \equiv B$ has charge -1/3;
\item
for the triplet of hypercharge $Y=-\frac{1}{3}$, $\rho^+ \equiv T$ has charge 2/3, $\rho^3 \equiv B$ has charge -1/3, 
and $\rho^- \equiv Y$ has charge -4/3,
\end{itemize}
as we expect.
\subsubsection*{Neutral couplings: the Z boson}
The couplings of the quarks in the vector triplet to the $Z$ boson
are
\bea
\label{eq:ZcouplingsTri}
\lag_Z^{XTB}  &=&
	 \frac{g}{2 c_W}  Z_\mu \left[
		\overline{X} \gamma^\mu X \left(I^3_q- 2 s_W^2 Q_X\right) + 
		\overline{T} \gamma^\mu T \left(-2 s_W^2 Q_T\right) +
		\overline{B} \gamma^\mu B  \left(-I_3^q- 2 s_W^2 Q_B\right) \right] \nn
	&=& \frac{g}{2 c_W} Z_\mu \left[
		\overline{X} \gamma^\mu X \left(2-\frac{10}{3} s_W^2\right) + 
		\overline{T} \gamma^\mu T \left(-\frac{4}{3} s_W^2\right) +
		\overline{B} \gamma^\mu B  \left(-2+\frac{2}{3} s_W^2\right) \right] \; ;\nn
\nn
\lag_Z^{TBY}  &=&
	 \frac{g}{2 c_W}  Z_\mu \left[
		\overline{T} \gamma^\mu T \left(I^3_q- 2 s_W^2 Q_T\right) + 
		\overline{B} \gamma^\mu B \left(-2 s_W^2 Q_B\right) +
		\overline{Y} \gamma^\mu Y  \left(-I_3^q- 2 s_W^2 Q_Y\right) \right] \nn
	&=& \frac{g}{2 c_W} Z_\mu \left[
		\overline{T} \gamma^\mu T \left(2-\frac{4}{3} s_W^2\right) + 
		\overline{B} \gamma^\mu B \left(\frac{2}{3} s_W^2\right) +
		\overline{Y} \gamma^\mu T  \left(-2+\frac{8}{3} s_W^2\right) \right] \; 	.
\eea
%
%%%%\bea
%%%%	\lag_Z  &=&
%%%%	 \frac{g}{2 c_W}  Z_\mu \left[
%%%%		\overline{X} \gamma^\mu X \left(T^3_q-s_W^2 Q_X\right) + 
%%%%		\overline{T} \gamma^\mu T \left(- s_W^2 Q_T\right) +
%%%%		\overline{B} \gamma^\mu B  \left(-T_3^q- s_W^2 Q_B\right) \right] \nn
%%%%	&=& \frac{g}{2 c_W} Z_\mu \left[
%%%%		\overline{X} \gamma^\mu X \left(1-\frac{5}{3} s_W^2\right) + 
%%%%		\overline{T} \gamma^\mu T \left(-\frac{2}{3} s_W^2\right) +
%%%%		\overline{B} \gamma^\mu B  \left(-1+\frac{1}{3} s_W^2\right) \right] \;.
%%%%\eea
This is perfectly consistent with what one expects from the Standard Model case 
(hypercharge minus twice the electric charge).
\subsubsection*{Charged couplings}
The couplings to the charged gauge bosons are 
\bea
	\label{eq:CCcouplingsTri}
	\lag_W^{XTB} &=& g W_\mu^+ \left( 
			\overline{T} \gamma^\mu B -  \overline{X} \gamma^\mu T \right) 
			+ g W_\mu^- \left( 
			\overline{B} \gamma^\mu T -  \overline{T} \gamma^\mu X \right)
\; ,\nn\nn
	\lag_W^{TBY} &=& g W_\mu^+ \left( 
			\overline{T} \gamma^\mu B -  \overline{X} \gamma^\mu T \right) 
			+ g W_\mu^- \left( 
			\overline{B} \gamma^\mu T -  \overline{T} \gamma^\mu X \right)
\eea
and similarly for the $Y=-\frac{1}{3}$ triplet. Notice that these couplings are 
a factor $\sqrt{2}$ larger than the Standard Model couplings (eq.~(\ref{eq:couplingsSM})).

%\subsection*{RAMARKS on the normalization of the neutral current couplings}
%%%In the results I presented here I adopted the convention
%%%for which the $W_\mu^a$ gauge bosons are multiplied by $\frac{\sigma^a}{2}$ 
%%%and the $U(1)$ generator is $\frac{1\!\!1}{2}$.  
%%%In the conventions followed by many others (also Heyenmaeyer and Lavoura/Silva
%%%in giving the analytical expressions) the  $U(1)$ generator is $1\!\!1$. 
%%%The charged current couplings are not affected, and for the neutral couplings one has 
%%%%
%%%\
%%%For the vector triplet, 
%%%\bea
%%%	\label{eq:NCcouplingsTri}
%%%	\lag_\gamma'  &=&  e \gamma_\mu \left[
%%%		\bar{\rho}^+ \gamma^\mu \rho^+ (Y+\frac{T^3_q}{2}) + \bar{\rho}^3 \gamma^\mu \rho^3 (Y) + \bar{\rho}^- \gamma^\mu \rho^- (Y-Y+\frac{T^3_q}{2}) \right] \;, \nn
%%%%
%%%\lag_Z'  &=&
%%%	 \frac{g}{2 c_W}  Z_\mu \left[
%%%		\overline{X} \gamma^\mu X \left(T^3_q- 2 s_W^2 Q_X\right) + 
%%%		\overline{T} \gamma^\mu T \left(-2 s_W^2 Q_T\right) +
%%%		\overline{B} \gamma^\mu B  \left(-T_3^q- 2 s_W^2 Q_B\right) \right] \nn
%%%	&=& \frac{g}{2 c_W} Z_\mu \left[
%%%		\overline{X} \gamma^\mu X \left(2-\frac{10}{3} s_W^2\right) + 
%%%		\overline{T} \gamma^\mu T \left(-\frac{4}{3} s_W^2\right) +
%%%		\overline{B} \gamma^\mu B  \left(-2+\frac{2}{3} s_W^2\right) \right] \;,
%%%\eea
%%%and
%%%\beq
%%%	T'^{3}_q = 2 \qquad,\quad Y' = \frac{2}{3} \;.
%%%\eeq
%As one sees, this factor $\frac{1}{2}$ in front of the $U(1)$ generator has 
%only the effect of changing the normalization of $T^3_q$, as expected. In 
%the calculation I stick to the convention that has the Standard Model $T^3$ to
%one.
%
\subsection{Physical couplings}
To obtain the physical fields we follow the procedure described in 
eqs.~(\ref{eq:gauge_eigenst_fields} - \ref{eq:rotation_matrices})
%%\beq
%%	\chi_{L,R}^q = \left(\begin{matrix} q_{L,R}\\Q_{L,R}\end{matrix}\right)
%%	=
%%	V_{L,R}^q \left(\begin{matrix} q^0_{L,R}\\Q^0_{L,R}\end{matrix}\right) \;,
%%\eeq
%%where $q=\{t, b\}$ and $Q=\{T, B\}$ and the rotation matrices are
%%%
%%\beq
%%	V_{L,R}^{q}=\left(
%%		\begin{matrix}
%%			c_{L,R}^{q}& -s_{L,R}^{q}\\
%%			s_{L,R}^{q} & c_{L,R}^{q}
%%		\end{matrix}
%%	\right)\; , 
%%	\qquad 
%%	\begin{array}
%%	cc_{L,R}^q  = \cos\theta_{L,R}^q \\ s_{L,R}^q = \sin\theta_{L,R}^q 
%%	\end{array} \;.
%%\eeq
%%Then the diagonal mass matrices read
%%\beq
%%	M^q_{diag} = V_L^q M^{q}_0 \left(V_R^q\right)^\dagger =
%%	\left(\begin{matrix}m_q & 0 \\0 &m_Q \end{matrix}\right) 
%%	 \;.
%%\eeq
\subsubsection*{$Y=\frac{2}{3}$ triplet $(X,T,B)$}
The physical couplings to the electroweak gauge bosons read
(cfr. eqs.~(\ref{eq:couplingsSM}), (\ref{eq:ZcouplingsTri}), (\ref{eq:CCcouplingsTri})) 
\beq
	\begin{array}{lclclclc}
	X_{XX,L} & = & \left( \begin{matrix} 2 \end{matrix}\right) &\;, \quad &
	X_{XX,R} &= & \left( \begin{matrix} 2 \end{matrix}\right) & \, , \nn
	X_{tt,L} & = & \left( \begin{matrix} (c_L^t)^2 & c_L^t s_L^t \\
									c_L^t s_L^t & (s_L^t)^2 \end{matrix}\right) & \;, \quad &
	X_{tt,R} & = & \left( \begin{matrix} 0 & 0 \\ 0 & 0 \end{matrix}\right) & \;, \nn
	X_{bb,L} & = & \left( \begin{matrix} -1-(s_L^b)^2 & c_L^b s_L^b \\
									c_L^b s_L^b & -1-(c_L^b)^2 \end{matrix}\right) & \;, \quad &
	X_{bb,R} & = & \left( \begin{matrix} -2 (s_R^b)^2 & 2 c_R^b s_R^b \\ 
							 2 c_R^b s_R^b  & -2 (c_R^b)^2 \end{matrix}\right) & \;, \nn
	A_{Xt,L} & = & \left( \begin{matrix} -\sqrt{2} s_L^t & \sqrt{2} c_L^t \end{matrix}\right) & \;, \quad &
	A_{Xt,R} & = & \left( \begin{matrix} -\sqrt{2} s_R^t & \sqrt{2} c_R^t \end{matrix}\right) & \;, 
	 \nn
	A_{tb,L} & = & \left( \begin{matrix} 
		c_L^b c_L^t + \sqrt{2} s_L^b s_L^t & c_L^t s_L^b - \sqrt{2} c_L^b s_L^t \\
		c_L^b s_L^t - \sqrt{2} c_L^t s_L^b  & s_L^b s_L^t + \sqrt{2} c_L^b c_L^t
	\end{matrix}\right) & \;, \quad &
	A_{tb,R} & = & \left( \begin{matrix} 
		\sqrt{2} s_R^b s_R^t & - \sqrt{2} c_R^b s_R^t \\
		 - \sqrt{2} c_R^t s_R^b  & \sqrt{2} c_R^b c_R^t
	\end{matrix}\right) & \;.
	\end{array}
\eeq
This is in agreement with the results of Ref.~\citep{Aguilar-Saavedra:2013qpa}. 

We also show here how one derives the relations among masses and angles 
of eq.~(\ref{rels}) for the $(X,T,B)$ triplet.
%%%Finally, the allowed mass terms (i.e., $SU(2)_L \times U(1)_Y$--invariant) that 
%%%couple the triplet $\rho$ to the Standard Model quarks are\footnote{
%%%To write down the mass Lagrangian, one just needs to notice that the $\{XTB\}$
%%%triplet has the same hypercharge as  $t_R$, while the $\{TBY\}$ has the same as 
%%%$b_R$. Therefore their coupling with the Higgs doublet is written goes exactly 
%%%as the SM couplings}\footnote{
%%%We initially defined $\rho = \rho^a\frac{\sigma^2}{2}$. For ease of comparison 
%%%with~\cite{Aguilar-Saavedra:2013qpa}, we absorbe the factor of 2 at the denominator 
%%%in the couplings $\lambda_7, \lambda_8$. }
%%%\beq
%%%	-{\cal L}_{XTB} = \lambda_7 \bar{\psi}_L \sigma^a \rho^a \tilde{H} 
%%%	+ M_{XTB} \bar{\rho} \rho \;,
%%%\eeq
Starting from the Lagrangian~(\ref{genpot}), the bare mass matrices are
\beq
	M^t = \left(\begin{matrix}
		\lambda_t \frac{v}{\sqrt{2}} & \lambda_7 \frac{v}{\sqrt{2}} \\
		0 & M_{XTB}	\end{matrix}\right) \qquad,\quad
		M^b = \left(\begin{matrix}
		\lambda_b \frac{v}{\sqrt{2}} & \lambda_7 v \\
		0 & M_{XTB}	\end{matrix}\right) \;.
\eeq
Let us notice that 
\beq
\label{eq:MphyssqFromM0}
	\left(M^q_{diag}\right)^2 = V_L^q M^{q} M^{q \dagger}  V_L^{q,\dagger} =
  V_R^q M^{q \dagger} M^{q}  V_R^{q \dagger} \qquad(q=t,b)\;.
 \eeq
The condition 
$$\left(M^q_{diag}\right)^2_{(1,1)}=m_q^2$$
 yields
\beq
\label{eq:triplet:LRangle_rels}
	\tan \theta_R^q = \frac{m_q}{M_Q} \tan\theta_L^q \;.
\eeq 
Next, one can ``reconstruct" the square bare mass matrix, both in the top and 
bottom sector, inverting eq.~(\ref{eq:MphyssqFromM0}). Imposing that the 
entry $(2,2)$ is the same and equals $M_{XTB}^2 = M_X^2$, we get
\beq
\label{eq:XTB_M_relations}
	M_X^2 = (c_{L,R}^b)^2 M_B^2+ (s_{L,R}^b)^2 m_b^2 = 
	(c_{L,R}^t)^2 M_T^2+ (s_{L,R}^t)^2 m_t^2 \;.
\eeq
%We will use these relations in particular for the left sector. 
%%Similarly, the relation
%%$$ V_L^{b \dagger} (M_{diag}^b)^2 V_L^b = M^b M^{b \dagger}
%%\equiv \sqrt{2} V_L^{t \dagger} (M_{diag}^t)^2 V_L^t  $$
Finally, using 
\beq
	 V_L^{q \dagger} (M_{diag}^q)^2 V_L^q = M^q M^{q \dagger} \qquad (q=t,b) \;,
\eeq
and noticing that the entries $(1,2)$ of these matrices are related,
\beq
	\left[M^b M^{b \dagger}\right]_{(1,2)} = \lambda_7 M_{XTB}
	=\sqrt{2} \left[M^t M^{t \dagger}\right]_{(1,2)}
\eeq
yields
\beq
\label{eq:LbvsLt_XTB}
	 (M_B^2-m_b^2) \sin 2 \theta_L^b = \sqrt{2} (M_T^2-m_t^2) \sin 2 \theta_L^t  \;.	
\eeq
\subsubsection*{$Y=-\frac{1}{3}$ triplet $(T,B,Y)$}
The physical couplings to the electroweak gauge bosons read
(cfr. eqs.~(\ref{eq:couplingsSM}), (\ref{eq:ZcouplingsTri}), (\ref{eq:CCcouplingsTri})) 
\beq
	\begin{array}{lclclclc}
	X_{tt,L} & = & \left( \begin{matrix} 1+(s_L^t)^2 & -c_L^t s_L^t \\
									-c_L^t s_L^t & 1+(c_L^t)^2 \end{matrix}\right) & \;, \quad &
	X_{tt,R} & = &  \left( \begin{matrix} 2 (s_R^t)^2 & -2 c_R^t s_R^t \\ 
							 -2 c_R^t s_R^t  & 2 (c_R^t)^2 \end{matrix}\right)& \;, \nn
	X_{bb,L} & = & \left( \begin{matrix} -(c_L^b)^2 & -c_L^b s_L^b \\
									-c_L^b s_L^b & -(s_L^b)^2 \end{matrix}\right) & \;, \quad &
	X_{bb,R} & = & \left( \begin{matrix} 0 & 0 \\ 0 & 0 \end{matrix}\right) & \;, \nn
	X_{YY,L} & = & \left( \begin{matrix} -2 \end{matrix}\right) &\;, \quad &
	X_{YY,R} &= & \left( \begin{matrix} -2 \end{matrix}\right) & \, , \nn
	A_{tb,L} & = & \left( \begin{matrix} 
		c_L^b c_L^t + \sqrt{2} s_L^b s_L^t & c_L^t s_L^b - \sqrt{2} c_L^b s_L^t \\
		c_L^b s_L^t - \sqrt{2} c_L^t s_L^b  & s_L^b s_L^t + \sqrt{2} c_L^b c_L^t
	\end{matrix}\right) & \;, \quad &
	A_{tb,R} & = & \left( \begin{matrix} 
		\sqrt{2} s_R^b s_R^t & - \sqrt{2} c_R^b s_R^t \\
		 - \sqrt{2} c_R^t s_R^b  & \sqrt{2} c_R^b c_R^t 
	\end{matrix}\right) & \;, \nn
 	A_{bY,L} & = & \left( \begin{matrix} -\sqrt{2} s_L^b \\ \sqrt{2} c_L^b \end{matrix}\right) & \;, \quad &
	A_{bY,R} & = & \left( \begin{matrix} -\sqrt{2} s_R^b \\ \sqrt{2} c_R^b \end{matrix}\right) & \; . %
	\end{array}
\eeq
%
%The mass Lagrangian is
%\beq
%	-{\cal L}_{TBY} = \lambda_8 \bar{\psi}_L \sigma^a \rho^a H 
%	+ M_{TBY} \bar{\rho} \rho \;,
%\eeq
%i.e.
The top and bottom mass matrices are (eq.~(\ref{genpot}))
\beq
	M^t = \left(\begin{matrix}
		\lambda_t \frac{v}{\sqrt{2}} & \lambda_8 v \\
		0 & M_{TBY}	\end{matrix}\right) \qquad,\quad
		M^b = \left(\begin{matrix}
		\lambda_b \frac{v}{\sqrt{2}} & \lambda_8 \frac{v}{\sqrt{2}} \\
		0 & M_{TBY}	\end{matrix}\right) \;.
\eeq
%For the relations among angles and masses, the proofs of 
Following the same proof that lead to eqs.~(\ref{eq:triplet:LRangle_rels}) 
and~(\ref{eq:XTB_M_relations}), we obtain 
\bea
	\tan \theta_R^q &=& \frac{m_q}{M_Q} \tan\theta_L^q \;, \nn
	M_Y^2 & = & (c_{L,R}^b)^2 M_B^2+ (s_{L,R}^b)^2 m_b^2 \nn
	& = &(c_{L,R}^t)^2 M_T^2+ (s_{L,R}^t)^2 m_t^2 \;.
\eea
Similarly, the equivalent of eq.~(\ref{eq:LbvsLt_XTB}) %, that came from 
%%$$
%%	(M_0^{t,XTB})_{(1,2)} = \frac{1}{\sqrt{2}} (M_0^{b,XTB})_{(1,2)} \;,
%%$$
is
\beq
		(M_T^2-m_t^2) \sin 2 \theta_L^t  = \sqrt{2} (M_B^2-m_b^2) \sin 2 \theta_L^b \;.	
\eeq

\section{EFT Coefficients and limits from $b$ and Higgs Couplings}
\label{eftapp}
Searches for VLQs at the LHC suggest that the masses are relatively heavy, 
\mbox{$M\gsim {\cal O}(800-1000)$~GeV}~\cite{ATLAS-CONF-2016-101,ATLAS-CONF-2016-102,
ATLAS-CONF-2016-013,Aad:2016shx,Aad:2016qpo,Aad:2015voa, 
Aad:2015kqa,Aad:2015mba,Khachatryan:2015gza,Khachatryan:2015axa,
Chatrchyan:2013wfa,Chatrchyan:2013uxa, ATLAS-CONF-2017-015}. 
This means 
that we are always in the regime,
\begin{equation}
{m_t\over M}<<1,\quad {m_b\over M}\sim 0, 
\end{equation}
where an effective field theory approach is warranted. The Lagrangian
involving third generation SM quarks and VLQs  can be written as,
\begin{eqnarray}
L&=&L_{Y,SM}+L_{KE}+L_{Q}\,
\end{eqnarray}
where $L_{Y,SM}$ is defined in Eq.~\ref{ysm}, $L_Q$ contains the VLQ
interactions given in Eq.~\ref{genpot} and $L_{KE}$ is the kinetic
energy term. 
At tree level, the heavy VLQs can be integrated out using the equations of 
motion~\cite{AguilarSaavedra:2009mx,Chen:2014xwa,Delaunay:2013iia},
generating an effective low-energy 
Lagrangian that only contains SM fields,
\begin{equation}
L_{eff}=L_{Y,SM} + L_{KE}^\prime+L_{qH}
\end{equation}
where $L_{KE}^\prime$ now includes only the SM quarks and 
\begin{equation}
L_{qH}=\Sigma_i C_i O_i + h.c.\, 
\label{eftdef}
\end{equation}
contains the effective interactions of the SM quarks with the gauge and Higgs boson
through higher dimensional operators (we restrict to dimension-6 operators). 
We have normalized the coefficients of these operators to 
be ${\cal {O}}\left({1\over M^2}\right)$.
The new Higgs-fermion dimension-6 operators are~\cite{delAguila:2000aa}:
\begin{eqnarray}
O_{Ht}&=& i(H^\dagger D_\mu H)({\overline t}_R\gamma^\mu t_R)\nonumber \\
O_{Hb}&=&i(H^\dagger D_\mu H)({\overline b}_R\gamma^\mu b_R)\nonumber \\
O_{Hq}&=&i(H^\dagger D_\mu H)({\overline \psi}_L\gamma^\mu \psi_L)\nonumber \\
O_{Hq}^{s}&=& i(H^\dagger \sigma^a D_\mu H)
({\overline \psi}_L\sigma^a\gamma^\mu \psi_L)\nonumber \\
O_{HY}^b&=&(H^\dagger H)({\overline \psi}_L H b_R)\nonumber \\
O_{HY}^t&=&(H^\dagger H)({\overline \psi}_L {\tilde H} t_R)\nonumber \\
O_{Htb}&=& ({\tilde H}^\dagger iD_\mu H) ({\overline t}_R\gamma^\mu b_R)\, .
\end{eqnarray}
To ${\cal{O}}({1\over M^2})$, the coefficients of Eq.~\ref{eftdef} are given in Table~\ref{tab:eftcoefs} in terms of the Yukawa
couplings and we assume the splitting between the  VLQ masses in a given representation are small,
corresponding to small mixing angles.  The different VLQ representations have 
quite different patterns for the coefficients~\cite{delAguila:2000aa, Chen:2014xwa,Batell:2012ca}.

\begin{table}
\begin{center}
\begin{tabular}{|c|c|c|c|c|c|c|c|}
\hline\hline
 & $C_{Ht}$& $C_{Hb}$  & $C_{Hq}$  &  $C^s_{Hq}$  &  $C_{HY}^b$  & $C_{HY}^t$ &  $C_{Htb}$\\
 \hline\hline 
 $T$ & 0& 0& ${\lambda_1^2\over 4 M_T^2}$ & $ -{\lambda_1^2\over 4 M_T^2}$ & 0 &  ${\lambda_t\lambda_1^2\over 2 M_T^2}$ 
 &  0 \\
\hline
 $B$ & 0& 0& 
 $-{\lambda_2^2\over 4 M_B^2}$ & 
 $- {\lambda_2^2\over 4 M_B^2}$ & 
 ${\lambda_b\lambda_2^2\over 2 M_B^2}$ &  0 &
    0
 \\
 \hline
 $(T, B)$ & 
 $-{\lambda_4^2\over 2 M_T^2}$ & 
 $ {\lambda_5^2\over 2 M_T^2}$&0&0 & 
 ${\lambda_b\lambda_5^2\over 2 M_T^2}$ &  
 ${\lambda_t\lambda_4^2\over 2 M_T^2}$ & 
  ${\lambda_4\lambda_5\over M_T^2}$ \\
 \hline
 $(X,T)$ & 
 ${\lambda_3^2\over 2 M_X^2}$ & 
 0&0&0
 &0&${\lambda_t\lambda_3^2\over 2 M_X^2}$&
  0\\
 \hline
 $(B,Y)$ & 0 &
 $-{\lambda_{6}^2\over 2 M_B^2}$ & 
 0
 &0&
 ${\lambda_b\lambda_{6}^2\over 2 M_B^2}$ & 0 & 
  0\\
 \hline
 $(X,T,B)$&0&0&${3\lambda_7^2\over 4 M_T^2}$&
 ${\lambda_7^2\over 4 M_T^2}$
 &${\lambda_7^2\lambda_b\over  M_T^2}$&
 ${\lambda_7^2\lambda_t\over 2 M_T^2}$&
  0\\
 \hline
 $(T,B,Y)$ &0&0&
 $-{3\lambda_8^2\over 4 M_B^2}$&
 $-{\lambda_8^2\over 4 M_B^2}$&
 ${\lambda_8^2\lambda_b\over 2 M_B^2}$&
 ${\lambda_8^2\lambda_t\over M_B^2}$& 
  0\\
 \hline\hline
  \end{tabular}
 \caption{\label{tab:eftcoefs} EFT coefficients for VLQ models in the large VLQ mass limit.} 
 \end{center}
 \end{table}
 
 These operators generate non-SM interactions of the fermions with
 the gauge and Higgs bosons.  The interactions with the $W$ boson defined 
 in Eq.~\ref{wcoups} become in the EFT limit, 
 \begin{eqnarray}
 A_{tb}^L&=&1+v^2 C_{Hq}^s\nonumber\\
 A_{tb}^R&=& {v^2\over 2}C_{Htb}\, ,
 \end{eqnarray}
and the couplings to the $Z$ boson defined in Eq. \ref{zdef} are,
 \begin{eqnarray}
 \delta X_{tt}^L&=& -{v^2\over 2}(C_{Hq}-C_{Hq}^s)\nonumber \\
 \delta X_{tt}^R&=& -{v^2\over 2}  C_{Ht}\nonumber \\
\delta X_{bb}^L&=&-{v^2\over 2}(C_{Hq}+C_{Hq}^s)\nonumber \\
\delta X_{bb}^R&=& -{v^2\over 2}  C_{Hb}\, . 
\end{eqnarray}
From Table~\ref{tab:eftcoefs}, we see that right-handed $W$ couplings are only generated in the $(TB)$  model,
while non-standard $W_L$ couplings arise in the singlet and triplet models.  In a similar fashion,
the doublet VLQ models have SM couplings of the $Z$ boson to the top and bottom quarks. 
Measuring the 
gauge boson fermion couplings puts strong constraints on the possible VLQ representations.

Finally, the $t,b$ couplings to the Higgs boson are also 
modified,
\begin{eqnarray}
L_{Y,SM}&\rightarrow& L_h\equiv -Y_f {\overline f} f h \nonumber \\
Y_f&=&{1\over \sqrt{2}}\biggl(\lambda_f-{3 v^2 \over 2}C_{HY}^f\biggr)\, ,
\end{eqnarray}
corresponding  to,
\begin{equation}
{m_f\over v}=Y_f +{v^2\over \sqrt{2}}C_{HY}^f\, .
\end{equation}
For the singlet and triplet models, we have the interesting relation between the $Zf {\overline {f}}$ couplings and the Higgs Yukawa
coupling,
\begin{equation}
\lambda_f(\delta X^L_{ff}-2I_3^f\delta X^R_{ff})=-{v^2\over 2} C_{HY}^f\quad
{\hbox{singlet, triplet VLQs}}\, . 
\end{equation}
For non-zero $C_{HY}^f$, the Yukawa coupling is no longer proportional to the mass, leading to flavor non-diagonal Higgs-fermion
interactions~\cite{Harnik:2012pb,Altmannshofer:2015esa}.

\end{appendix}

%%%%%%%%%%%%%%%%%%%%%%%%%%%%%%%%%%%%%%%%%%%%%%%%%%%%%%%%%%%%%%%%%%%%%%%
\bibliographystyle{unsrt}
\bibliography{vlq}
%%%%%%%%%%%%%%%%%%%%%%%%%%%%%%%%%%%%%%%%%%%%%%%%%%%%%%%%%%%%%%%%%%%%%%%

\end{document}